\documentclass[article]{iacrtrans}

\usepackage{algorithmic}
\usepackage{graphicx}
\usepackage{textcomp}
\usepackage{xcolor}
\usepackage{fancyhdr}
\usepackage{tikz}
\usepackage{tikzsymbols}
\usetikzlibrary{positioning}
\usepackage{pgfplots}
\pgfplotsset{compat=1.18}
\usetikzlibrary{patterns}
\usepackage[boxruled,algo2e]{algorithm2e}
\usepackage{svg}

\usepackage{algorithm}
\usepackage{comment}
\usepackage{makecell}
\usepackage{booktabs}
\usepackage{multirow}
\usepackage{hyperref}
\usepackage{amsthm}

\usepackage[perpage]{footmisc}
\usepackage{adjustbox}
\hypersetup{breaklinks=true}
\usepackage{soul}

\newcommand{\papertitle}{REED\xspace}
\newcommand{\papertitlemain}{RE3D\xspace}



\newcommand{\bigo}[1]{\mathcal{O}({#1})}
\newcommand{\polring}[1]{\mathcal{R}_{#1}}

\usepackage{tikz}
\DeclareRobustCommand{\ballnumber}[1]{\tikz[baseline=(myanchor.base)] \node[circle,fill=.,inner sep=1pt] (myanchor) {\color{-.}\bfseries\footnotesize #1};}

\author{Aikata Aikata\inst{1} \and Ahmet Can Mert\inst{1} \and Sunmin Kwon\inst{2} \and Maxim Deryabin\inst{2} \and Sujoy Sinha Roy\inst{1}}
\institute{
  Graz University of Technology, Graz, Austria\\ \email{{aikata,ahmet.mert,sujoy.sinharoy}@tugraz.at}
  \and
  Samsung Advanced Institute of Technology, Samsung Electronics, Korea\\ \email{{sunmin7.kwon,max.deriabin}@samsung.com}
}

\title{REED: Chiplet-based Accelerator for Fully Homomorphic Encryption}

\begin{document}

\maketitle


\begin{abstract}
Fully Homomorphic Encryption (FHE) enables privacy-preserving computation and has many applications. However, its practical implementation faces massive computation and memory overheads. To address this bottleneck, several Application-Specific Integrated Circuit (ASIC) FHE accelerators have been proposed. All these prior works put every component needed for FHE onto one chip (monolithic), hence offering high performance. However, they encounter common challenges associated with large-scale chip design, such as inflexibility, low yield, and high manufacturing costs. In this paper, we present the \emph{first-of-its-kind} multi-chiplet-based FHE accelerator `\papertitle' for overcoming the limitations of prior monolithic designs.  To utilize the advantages of multi-chiplet structures while matching the performance of larger monolithic systems, we propose and implement several novel strategies in the context of FHE. These include a scalable chiplet design approach, an effective framework for workload distribution, a custom inter-chiplet communication strategy, and advanced pipelined Number Theoretic Transform and automorphism design to enhance performance.

Our instruction-set and power simulations experiments with a prelayout netlist indicate that \papertitle 2.5D microprocessor consumes 96.7mm$^2$ chip area, 49.4 W average power in 7nm technology. It could achieve a remarkable speedup of up to 2,991$\times$ compared to a CPU (24-core 2$\times$Intel X5690) and offer 1.9$\times$ better performance, along with a 50\% reduction in development costs when compared to state-of-the-art ASIC FHE accelerators. Furthermore, our work presents the \textit{first} instance of benchmarking an encrypted deep neural network (DNN) training. Overall, the \papertitle architecture design offers a highly effective solution for accelerating FHE, thereby significantly advancing the practicality and deployability of FHE  in real-world applications.
\end{abstract}

\section{Introduction}\label{sec:intro}

Data breaches can put millions of private accounts at risk because data is often stored or processed without encryption, making it vulnerable to attacks~\cite{breach1,breach2,breach3}. Fully Homomorphic Encryption (FHE) is a solution that allows secure, private computations, communications, and storage. It enables servers to compute on homomorphically encrypted data and return encrypted outputs. FHE has a wide range of applications, including cloud computing~\cite{ccs_appl1,ccs_appl3}, data processing~\cite{ccs_appl2}, and machine learning~\cite{eurosp_appl}. The concept of FHE was introduced in 1978 by Rivest, Adleman, and Dertouzos~\cite{RIVEST78}, and the first FHE scheme was constructed in 2009 by Gentry~\cite{GENTRY09}. Since then, many FHE schemes have emerged- BGV~\cite{DBLP:journals/eccc/BrakerskiGV11}, FV~\cite{DBLP:journals/iacr/FanV12}, CGGI~\cite{chillotti2020tfhe}, and CKKS~\cite{ckks_scheme,RNS_CKKS18,ccs_mk}. These schemes allow computations to be outsourced without the need to trust the service provider, providing a functional and dependable privacy layer.

Despite significant progress in the mathematical aspects of FHE, state-of-the-art FHE schemes typically introduce 10,000$\times$ to 100,000$\times$ slowdown~\cite{slowdown} compared to plaintext calculations. This overhead can be attributed to plaintext expanding into large polynomials when encrypted using an FHE scheme. Subsequently, simple operations, like plaintext multiplication, translate into complex polynomial operations. FHE's massive computation and data overhead hinders its deployment in real-life applications. To bridge this performance gap, researchers have proposed acceleration techniques on various platforms, including GPU, FPGA, and ASIC~\cite{FPT,wang_asic_gentry,cheetah,ARK,CraterLake,BTS_isca,basalisc,feldmann_2021f1,SHARP, takeshita2020, xin_iscas21,cofhee,tensorfhe,medha,heax,wang_fpga_gentry,roy_ches2015,JKACL21,badawi20,roy_hpca,roy_tc2018}. %
Software implementations offer flexibility but poor performance. Attempts have been made to provide GPU~\cite{JKACL21,badawi20} and FPGA-based solutions~\cite{medha,heax,roy_hpca}. However, the performance gap is still 2-3 orders compared to plain computation.

Currently, the fastest hardware acceleration results for FHE have been reported using ASIC modeling~\cite{ARK,CraterLake,BTS_isca,basalisc,feldmann_2021f1,SHARP}. The works propose utilizing large chip architecture designs with all FHE building blocks onto a single chip to maximize performance, hence monolithic. While simulations of these architectures show that they can achieve high performance for FHE workloads, the limitations of the current manufacturing capabilities, such as inflexibility, low yield, and higher manufacturing costs~\cite {chiplet_vs_mono}, impact their real-world deployment. For instance, the large architectures~\cite{BTS_isca, ARK, SHARP} with area-consumption of approximately 400mm$^2$, result in a manufacturing yield of only 67\%~\cite{chiplet_size_MaWWCH22}, chip fabrication cost of over 25 million US\$ \cite{pricing}, and long time-to-market ($>$3 years). 

Additionally, several of these proposals overlook the crucial need for communication-computation parallelism as the off-chip to on-chip communication is slower than the chip's computation speed. Our analysis shows that this feature is important in an FHE accelerator for achieving good performance when running complex tasks like neural network training. Prior works also utilize higher on-chip bandwidth due to readily available on-chip memory ($20$TB/s~\cite{ARK}, $36$TB/s~\cite{SHARP}, and $84$TB/s~\cite{CraterLake}). Replacing this on-chip memory with cheaper HBM3 ($1.2$TB/s bandwidth) would require 17 to 70 HBM3 modules to match the necessary bandwidth.

In summary, while the large and complex monolithic FHE architectures proposed in prior works show promise, they face practical challenges such as high manufacturing costs, yield rates, and extended time-to-market. Addressing these challenges opens the door to exploring new approaches like chiplet-based architecture design. Chiplet-based architecture design utilizes multiple smaller chiplets instead of one large monolithic chip to realize a large system. {Chiplets} are modular building blocks that are combined to create more complex integrated circuits, such as CPUs, GPUs, Systems-on-Chip (SoCs), or System-in-Package (SiPs).

The transition to chiplet integrated systems represents both the present and future of architectural designs~\cite{chiplet_vs_mono, manticore,  web_link1,chiplet_natalie,chiplet_size_GPG,chiplet_size_MaWWCH22}. In the DATE2024 keynote talk~\cite{web_link2}, the speaker remarks how chiplet-based designs help `push the performance boundaries, with maximum efficiency, while managing costs associated with manufacturing and yield'. {Chiplet-based architectures also feature the advantage of tiling beyond the reticle limit (858$mm^2$)~\cite{chiplet_size_GPG} as multiple chiplets can be integrated for better performance.} Although chiplet-based architectures enjoy the aforementioned advantages, they also face a trade-off between performance and yield. Multiple smaller chiplets offer high yields and reduced manufacturing costs but, at the same time, experience performance overhead due to slower chiplet-to-chiplet communication. Taking the advantages and challenges of chiplet-based systems into consideration, we are curious to investigate the following research questions:

\vspace{1em}

{\centering\textit{
How can we design and optimize a multi-chiplet accelerator for FHE that matches the performance of large monolithic FHE accelerators while overcoming the inherent challenges of monolithic designs?
} 
}

\vspace{1em}

To investigate the question mentioned above, we present \papertitle, a multi-chiplet architecture for FHE acceleration. We propose a holistic design methodology covering all aspects of FHE acceleration, from low-level building blocks to high protocol levels, and reduce the area to 43.9$mm^2$ for one \papertitle-chiplet. This includes the first scalable design methodology for one chiplet and ensures full utilization of chiplets for varying amounts of available off-chip data bandwidths. After finalizing an efficient design of one chiplet, we move to a data and task distribution study for multiple chiplets in the context of CKKS ~\cite{ckks_scheme} routines. Towards this, we contribute novel strategies that offer long-term computation and communication parallelism. Finally, we synthesize the proposed design methodology for ASIC and report application benchmarks.

\subsection*{Contributions} 
To the extent of our knowledge, this is the \textit{first chiplet-based architecture for accelerating FHE}. Throughout this work, we have followed Occam's razor, seeking the simplest solutions for the best results. We unfold our major contributions as follows:

\begin{itemize}

    \item \textbf{Chiplet-based FHE accelerator:} We present a novel and cost-effective chiplet-based FHE implementation approach, which is inherently scalable\footnote{Proof-of-concept implementation for multi-chiplet cycle accurate model, NTT/INTT unit, and Automorphism unit is open source and available at \url{https://github.com/aikata10/REED/}.}. The chiplets are homogeneous (i.e., identical), which reduces testing and integration costs. \papertitle with 2.5D packaging surpasses state-of-the-art work SHARP$_{64}$\cite{SHARP} with 1.9$\times$ better performance and 2$\times$ less development cost. 
    
    \item \textbf{Workload division strategy:}
    The first step to realizing a multi-chiplet architecture is to develop an efficient disintegration strategy that helps us divide the workloads among multiple chiplets and reduces memory consumption. Hence, we propose an interleaved data and workload distribution technique for all FHE routines.

    \item \textbf{FHE-tailored efficient C2C communication:} Chiplet-based architectures suffer from slow C2C (chiplet-to-chiplet) communication. We address this by proposing the \textit{first non-blocking ring-based inter-chiplet communication} strategy tailored to FHE. This mitigates data exchange overhead during the KeySwitch macro-routine, accelerating Bootstrapping (the most expensive FHE routine).
   
    \item \textbf{Scalable design:} To attain scalability by design, we propose a configuration-based design methodology such that the memory read/write and computational throughput are the same. Changing the configuration parameters allows the architecture to adapt to the desired area and throughput requirements. This also offers inherent communication-computation parallelism in the design of every chiplet.
    
    \item \textbf{Novel compute acceleration:} Furthermore, we present new design techniques for the micro-procedures of FHE- the number-theoretic transform (NTT) and automorphism (AUT). Our approach introduces Hybrid NTT, eliminating the need for expensive transpose operation and scratchpad memory. It is easily scalable for higher or lower polynomial degrees. Hence, other applications, such as zero-knowledge proofs, can also benefit from this, where transposition is expensive due to high polynomial degrees. Additionally, we have prototyped these building blocks on FPGA- AlveoU250.
    
    \item \textbf{Application benchmark:} Finally, we choose parameters offering high precision and good performance. \papertitle is the \textit{first work to benchmark an encrypted deep neural network training}, showcasing practical and real-world impact. While CPU (24-core, 2$\times$Intel Xeon CPU X5690 $@$ 3.47GHz) requires 29 days to finish it, \papertitle 2.5D would take only 15.4 minutes, a realistic time for an NN training. We also use DNN training to run accuracy/precision experiments and validate our parameter choice.
    
\end{itemize}

\subsection*{Connection and comparison with chiplet designs for ML}
    While prior chiplet-based Machine Learning (ML) works address similar problems, our solutions are tailored to meet FHE requirements more effectively. For instance, \cite{ml_chiplet} addresses MCM's long “tail-latency” issue using non-uniform work distribution and communication-aware data placement. In the context of FHE, we resolve this by running parallel computations over extended periods, ensuring uniform task distribution and data placement. Our chiplet interconnections are ring-like and unidirectional. 
    Although we do not propose an automatic tool, our analysis, similar to \cite{baton}, focuses on long-term chiplet utilization based on FHE's computational-depth. Our methodology introduces a new configuration-based design built from scratch with novel building blocks and high-level protocols. In contrast to \cite{centaur}, which combines heterogeneous-chiplets, we propose homogeneous-chiplets observing unique data-flow of FHE. A common limitation of the prior works is that they propose very small chiplet sizes (2 to 6 $mm^2$), which is too small as per a recent study done by the authors in~\cite{chiplet_size_GPG}. Thus, we ensure that our chiplet sizes fall within the optimal range.

\section{Background}\label{sec:background}
Let $\mathbb{Z}_Q$ represent the ring of integers in the $[0, Q-1]$ range. $\polring{{Q,N}}=\mathbb{Z}_Q[x]/(x^N+1)$ refers to polynomial ring containing polynomials of degree at most $N-1$ and coefficients in $\mathbb{Z}_Q$. In the Residue Number System (RNS)~\cite{rns} representation, $Q$ is a composite modulus comprising co-prime moduli, $Q=\prod_{i=0}^{L-1}q_i$.
The RNS representation is used to divide a big computation modulo $Q$ into much smaller computations modulo $q_i$ such that the small computations can be carried out in parallel.
With the application of RNS, a polynomial $a\in \polring{{Q,N}}$ becomes a vector, say $\boldsymbol{a}$, of residue polynomials. Let the $i$-th residue polynomial within $\boldsymbol{a}$ be denoted as $a^i \in \polring{{q_i,N}}$. We use the `monotype' font ($\mathtt{c}$/$\mathtt{sk}$) to represent ciphertexts/keys. Operators $\cdot$ and $\langle,\rangle$ denote the multiplication and dot-product between two ring elements. Noise ($e$) is refreshed for every computation. The tilde sign ($~\tilde{}~$) represents a data in NTT format (e.g., $\texttt{NTT}(a)$= $\tilde{a}$).

\subsection{FHE schemes and CKKS routines}
Different FHE schemes exist in literature, such as, BFV~\cite{DBLP:journals/iacr/FanV12}, BGV~\cite{DBLP:journals/eccc/BrakerskiGV11}, CGGI \cite{chillotti2020tfhe}, and CKKS~\cite{ckks_scheme,RNS_CKKS18}. These schemes use polynomial arithmetic but differ primarily in the data types they can encrypt. For instance, BGV and BFV encrypt integers, while CKKS encrypts fixed-point numbers. Due to the support for fixed-point arithmetic, CKKS is widely adopted for benchmarking machine learning applications~\cite{helr,KSKLC18}. Therefore, this work targets the RNS (Residue Number System) CKKS~\cite{RNS_CKKS18}. Other FHE schemes like BGV and B/FV are also based on RLWE and require similar operations as in CKKS. Thus, these schemes can utilize the same design methodology for varying parameters.
In the following, we briefly describe the main procedures within the RNS CKKS~\cite{RNS_CKKS18,KimPP22,HK20} for ciphertexts at level $l$ (multiplicative depth is $l$) where $l < L$, $Q_{l} = \prod_{i=0}^{l} q_i$, and $L$ is the maximum level. The residue polynomial associated with each modulus $q_i$ in the RNS representation is commonly called the RNS limb. A CKKS ciphertext consists of components, e.g., $\mathtt{c}=  ({\boldsymbol{c}_0,\boldsymbol{c}_1})$, where $\boldsymbol{c}_0$ and $\boldsymbol{c}_1$ are vectors of limbs. \autoref{tab:param_notation} describes the CKKS parameters, and algorithmic descriptions are provided for $dnum=L+1 (K=1)$.
\begin{table}[t]
    \centering
    \caption{CKKS Parameters}
    \label{tab:param_notation}
    \begin{tabular}{l|l}
    \hline
    \textbf{Parameter} & \textbf{Definition}   \\\hline  \hline
    $N,n~(\leq \frac{N}{2})$ & Polynomial size, maximum slots packed   \\
    $Q$, $q_i$ & Coefficient modulus, RNS bases $Q = \prod_{i=0}^{L}q_i$  \\
    $L,l$ & Multiplicative depth (\#RNS bases - 1)  $l<L$   \\
    $dnum$ & Number of digits in the switching key\\
    $P$, $p_i$ & Special modulus and its RNS base   \\
    $K~(=\lceil \frac{L+1}{dnum} \rceil)$ & Number of RNS bases for $P = \prod_{i=0}^{K-1}p_i$  \\
     $w$ & Word size ($\log{p_i},\log{q_i}$)  \\
    $L_{\textit{boot}},L_{\textit{eff}}$ & Multiplicative depth of/after bootstrapping  \\
    \hline
    \end{tabular}
\end{table}

\begin{enumerate}
    \item $\mathtt{CKKS.Add}(\mathtt{c}, \mathtt{c}')$: As shown in \autoref{algo:add_14}, this operation takes two input ciphertexts $\mathtt{c} \text{ and } \mathtt{c}' $ and computes $\mathtt{c}_{\text{add}} = (\boldsymbol{d}_0, \boldsymbol{d}_1) = (\boldsymbol{c}_0 + \boldsymbol{c}'_0 , \boldsymbol{c}_1 + \boldsymbol{c}'_1)$.
    
    \item $\mathtt{CKKS.Mult}(\mathtt{c}, \mathtt{c}')$: It multiplies the two input ciphertexts $\mathtt{c}$ and $\mathtt{c}'$, as shown in \autoref{algo:mult_14}, and computes the non-linear ciphertext $\mathtt{d}=(\boldsymbol{d}_0, \boldsymbol{d}_1, \boldsymbol{d}_2)$ = ($ {\boldsymbol{c}_0\cdot \boldsymbol{c}'_0}, {\boldsymbol{c}_0\cdot \boldsymbol{c}'_1} + {\boldsymbol{c}_1 \cdot \boldsymbol{c}'_0}, {\boldsymbol{c}_1 \cdot \boldsymbol{c}'_1} $). Subsequently, $\mathtt{CKKS.KeySwitch}$ transforms $\mathtt{d}$ into a linear ciphertext. The computation is done on data in NTT format.
    
    \vspace{-0.55cm}
    \noindent\begin{minipage}[t]{0.45\textwidth}
    \begin{algorithm}[H]
    \renewcommand{\algorithmicrequire}{\textbf{In:}}
    \renewcommand{\algorithmicensure}{\textbf{Out:}}
    \caption{\texttt{CKKS.Add} ~\cite{RNS_CKKS18}}
    \label{algo:add_14}
        \begin{flushleft}
\textbf{In:}  $\mathtt{c}=(\mathbf{\tilde{c}}_{0}, \mathbf{\tilde{c}}_{1})$, $\mathtt{c}'=(\mathbf{\tilde{c}}'_{0}, \mathbf{\tilde{c}}'_{1}) \in R_{Q_l}^2$

 \textbf{Out:} $\mathtt{c}_{\text{add}}=(\mathbf{\tilde{d}}_{0}, \mathbf{\tilde{d}}_{1}) \in R_{Q_l}^{2}$
\end{flushleft}
\vspace{-12pt}
    \begin{algorithmic}[1]
    	\STATE $\mathbf{\tilde{d}}_{0} \leftarrow \mathbf{\tilde{c}}_{0} + \mathbf{\tilde{c}}'_{0}$, $\mathbf{\tilde{d}}_{1} \leftarrow \mathbf{\tilde{c}}_{1} + \mathbf{\tilde{c}}'_{1}$
    \end{algorithmic}
    \end{algorithm}
    \vspace{-24pt}
    \begin{algorithm}[H]
    \renewcommand{\algorithmicrequire}{\textbf{In:}}
    \renewcommand{\algorithmicensure}{\textbf{Out:}}
    \caption{\texttt{CKKS.Rotate} ~\cite{RNS_CKKS18}}
    \label{algo:rot_14}
        \begin{flushleft}
\textbf{In:}  $\mathtt{c}=(\mathbf{\tilde{c}}_{0}, \mathbf{\tilde{c}}_{1}) \in R_{Q_l}^2$, $rot$

 \textbf{Out:} $\mathtt{c}_{\text{rot}}=(\mathbf{\tilde{d}}_{0}, \mathbf{\tilde{d}}_{1}) \in R_{Q_l}^{2}$
\end{flushleft}
\vspace{-12pt}
    \begin{algorithmic}[1]
    	\STATE $ (\mathbf{\tilde{d}}_{0},\mathbf{\tilde{d}}_{1}) \leftarrow \rho_{rot}(\mathbf{\tilde{c}}_{0},\mathbf{\tilde{c}}_{1})$
    \end{algorithmic}
    \end{algorithm}
\end{minipage}
\hfill
\begin{minipage}[t]{0.45\textwidth}
    \vspace{0pt}
    \begin{algorithm}[H]
    \renewcommand{\algorithmicrequire}{\textbf{In:}}
    \renewcommand{\algorithmicensure}{\textbf{Out:}}
    \caption{$\mathtt{CKKS.Mult}$ ~\cite{RNS_CKKS18}}
    \label{algo:mult_14}
    \begin{flushleft}
\textbf{In:}  $\mathtt{ct}=(\mathbf{\tilde{c}}_{0}, \mathbf{\tilde{c}}_{1}) \in R_{Q_l}^2$

\textbf{In:}  $\mathtt{ct}'=(\mathbf{\tilde{c}}'_{0}, \mathbf{\tilde{c}}'_{1}) \in R_{Q_l}^2$

 \textbf{Out:} $\mathtt{d}=(\mathbf{\tilde{d}}_{0}, \mathbf{\tilde{d}}_{1}, \mathbf{\tilde{d}}_{2}) \in R_{Q_l}^{3}$ 
\end{flushleft}
\vspace{0.5pt}
    \begin{algorithmic}[1] 
    	\STATE $\mathbf{\tilde{d}}_{0} \leftarrow \mathbf{\tilde{c}}_{0} \cdot \mathbf{\tilde{c}}'_{0}$
     \STATE $\mathbf{\tilde{d}}_{2} \leftarrow \mathbf{\tilde{c}}_{1} \cdot \mathbf{\tilde{c}}'_{1}$
    	\STATE $\mathbf{\tilde{d}}_{1} \leftarrow \mathbf{\tilde{c}}_{0} \cdot \mathbf{\tilde{c}}'_{1}$
     \STATE $\mathbf{\tilde{d}}_{1} \leftarrow \mathbf{\tilde{d}}_{1}  + \mathbf{\tilde{c}}_{1} \cdot \mathbf{\tilde{c}}'_{0}$
    \end{algorithmic}
    \end{algorithm}
\end{minipage}

    \item $\mathtt{CKKS.Rotate}(\mathtt{c},\textit{rot}, \mathtt{{ksk}_{rot}})$:
It rotates the plaintext slots within $\mathtt{c}$ by $\textit{rot}$. First, a permutation $\rho$ is applied to the ciphertext polynomial coefficients, as shown in \autoref{algo:rot_14}. This permutation is called automorphism and is determined by the Galois element $\textit{gle}=5^{\textit{rot}}\bmod{2N}$. Finally, the permuted ciphertext is processed by $\mathtt{CKKS.KeySwitch}$ using the rotation key $\mathtt{{ksk}_{rot}}$.

    \item $\mathtt{CKKS.KeySwitch}(\mathtt{d},\mathtt{{ksk}})$: It uses a KeySwitch or evaluation key $\mathtt{ksk}$ to homomorphically transform a ciphertext decryptable under one key into a new ciphertext decryptable under another (original) key, as illustrated in \autoref{algo:relin_14}. It computes $\mathtt{c}''$ where ${\boldsymbol{c}_0''} = \sum^{l-1}_{i=0} {d}_{2}^{i}\cdot {{ksk}}_{0}^{i} \in \polring{PQ_l, N}$ and ${\boldsymbol{c}_1''} = \sum^{l-1}_{i=0} {d}_{2}^{i} \cdot {{ksk}}_{1}^{i} \in \polring{PQ_l, N}$. This is followed by $\mathtt{c} = \big((d_0, d_1) + \mathtt{CKKS.ModDown}(\mathtt{c''})\big) \in\polring{Q_l, N}^2$. $\mathtt{CKKS.ModDown}()$  scales down the modulus (${PQ_l}$ to ${Q_l}$), and is described in \autoref{algo:moddown_14} following the works~\cite{HK20,KimPP22}. A more detailed and generalized description is provided in \autoref{algo:relin_gen} (\autoref{appendix:keyswitch}).

    \item $\mathtt{CKKS.Bootstrap}$: It refreshes a noisy ciphertext~\cite{bootstrap1, bootstrap2, bootstrap3} by producing a new ciphertext with a higher depth or lower noise. As bootstrapping itself consumes a certain number of depths, the depth of a bootstrapped ciphertext, say $L_{\textit{eff}}$, is smaller than the initial depth $L$ after fresh encryption. Bootstrapping is required to refresh the processed ciphertexts in complex applications, such as DNN. It consists of the following four major steps.

    \begin{itemize}
    \item   \textbf{Slot to Coefficient Conversion:} Converts the ciphertext from slot form to polynomial form using a homomorphic DFT (Discrete Fourier Transformation). This involves computing a homomorphic matrix-vector multiplication, where the matrix is not encrypted, and the ciphertext is the vector.
        \item \textbf{Modulus Raising:}  Raises the modulus from $q_0$ to $Q$, introducing an error term that needs removal. This requires a plain ModUp operation,
        \item \textbf{Coefficient to Slot Conversion:} Converts the ciphertext back to slot form after modulus raising via homomorphic IDFT (Inverse DFT) computation. This also involves computing a homomorphic matrix-vector multiplication.
        \item \textbf{Homomorphic Modular Reduction:} This step applies a Chebyshev polynomial approximation to remove the introduced error term after Modulus raising. It results in refreshed computational depth. 
    \end{itemize}
\begin{minipage}[t]{0.85\textwidth}
\vspace{-20pt}
    \begin{algorithm}[H]
\renewcommand{\algorithmicrequire}{\textbf{In:}}
\renewcommand{\algorithmicensure}{\textbf{Out:}}
\caption{$\mathtt{CKKS.KeySwitch}$ ~\cite{HK20,KimPP22} (for $dnum=L+1$)}
\label{algo:relin_14}
\begin{flushleft}
\textbf{In:}  $\mathtt{d} = (\mathbf{\tilde{{d}}}_{0}, \mathbf{\tilde{{d}}}_{1}, \mathbf{\tilde{{d}}}_{2}) \in R_{Q_l}^{3}$, $\tilde{\mathtt{ksk}}_0 \in R_{PQ_l}^l, \tilde{\mathtt{ksk}}_1 \in R_{PQ_l}^l$ 

 \textbf{Out:} $\mathtt{d}' = (\mathbf{\tilde{{d}}}'_{0}, \mathbf{\tilde{{d}}}'_{1}) \in R_{Q_l}^{2}$ 
\end{flushleft}
\vspace{-10pt}
\begin{algorithmic}[1]
    \FOR{$j=0$ to $l$}	
	\STATE ${d}_{2}[j] \leftarrow \mathtt{INTT}({\tilde{{d}}}_{2}[j]) \in \mathbb{Z}_{q_j}$
    \ENDFOR
    \FOR{$j=0$ to $l+1$}	
	    \STATE $(\mathbf{\tilde{{c}}}''_{0}[j], \mathbf{\tilde{{c}}}''_{1}[j]) \leftarrow 0$
	    \FOR{$i=0$ to $l$}	
    	\STATE ${\tilde{{r}}} \leftarrow \mathtt{NTT}(\big[{d}_{2}[i]\big]_{q_j}) \in \mathbb{Z}_{q_j}$
    	\STATE ${\tilde{{c}}}''_{0}[j] \leftarrow \big[{\tilde{{c}}}''_{0}[j] + \tilde{\mathtt{ksk}}_{0}[i][j] \cdot {\tilde{{r}}}\big]_{q_j}$, ${\tilde{{c}}}''_{1}[j] \leftarrow \big[{\tilde{{c}}}''_{1}[j] + \tilde{\mathtt{ksk}}_{1}[i][j] \cdot {\tilde{{r}}}\big]_{q_j}$  
\ENDFOR	
	\ENDFOR	
    \STATE $\mathbf{\tilde{{d}}}'_{0} \leftarrow \mathbf{\tilde{{d}}}_{0} + \mathtt{CKKS.ModDown}(\mathbf{\tilde{{c}}}''_{0})$, $\mathbf{\tilde{{d}}}'_{1} \leftarrow \mathbf{\tilde{{d}}}_{1} + \mathtt{CKKS.ModDown}(\mathbf{\tilde{{c}}}''_{1})$   
\end{algorithmic}
\end{algorithm} 
\end{minipage} 

\begin{minipage}[t]{0.85\textwidth}
\vspace{-10pt}
    \begin{algorithm}[H]
    \renewcommand{\algorithmicrequire}{\textbf{In:}}
    \renewcommand{\algorithmicensure}{\textbf{Out:}}
    \caption{$\mathtt{CKKS.ModDown}$ ~\cite{RNS_CKKS18}}
    \label{algo:moddown_14}
    \begin{flushleft}
\textbf{In:} $\mathbf{\tilde{d}} \in R_{PQ_l}$ 

 \textbf{Out:} $\mathbf{\tilde{d}'} \in R_{Q_l}$ 
\end{flushleft}
\vspace{-10pt}
    \begin{algorithmic}[1] 
        \STATE $t \leftarrow \mathtt{INTT}({\tilde{d}}[l+1])$       
        \FOR{$i=0$ to $l$}	
    	\STATE $\tilde{t} \leftarrow \mathtt{NTT}(\big[{t}\big]_{q_i} \in \mathbb{Z}_{q_i}$
    	\STATE $\tilde{{d}'}[i] \leftarrow \big[p_0^{-1} \cdot (\tilde{{d}}[i] - \tilde{{t}})\big]_{q_i}$   
        \ENDFOR
    \end{algorithmic}
    \end{algorithm} 
\end{minipage}

\end{enumerate}

\subsection{FHE Hardware design goals}

A tiered structure exists in the CKKS scheme routines. The high-level or \emph{macro} rotuines are $\mathtt{CKKS.Add}$, $\mathtt{CKKS.Mult}$, $\mathtt{CKKS.Rotate}$, and $\mathtt{CKKS.KeySwitch}$.  These macro procedures apply micro procedures, such as forward and inverse Number Theoretic Transforms (NTT/INTT), dyadic Multiplication/Addition/Subtraction (MAS), and Automorphism (AUT). The NTT is used for multiplying two $N$ coefficients long polynomials in $\bigo{N\log{N}}$ time complexity, which is the asymptotically fastest one. 

The special $\mathtt{CKKS.Bootstrap}$ procedure uses these macro procedures in a specific sequence to refresh noisy ciphertexts.
Note that contrary to schemes like FHEW, TFHE \cite{chillotti2020tfhe} where bootstrapping is a standalone procedure, CKKS-Bootstrapping is a high-level routine which utilizes KeySwitches, Automorphisms, and MACs. Therefore, while TFHE/FHEW accelerators (e.g., \cite{FPT}) focus on optimizing the programmable bootstrapping, acceleration for CKKS relies on optimized KeySwitching, Automorphisms, and MACs. Among these operations, MAC is a straightforward linear operation. Automorphism involves permutation, followed by KeySwitch~\cite{HK20,KimPP22}. This permutation, if naively implemented, can become complex and expensive as the input polynomial has N=$2^{16}$ coefficients and offers N/2 different permutations. In this work, we show how we design the permutation unit that is not only cheap in terms of area but also has linear time complexity for all permutations. The final operation- KeySwitch, is the most expensive due to the expensive ModUp step (\autoref{fig:ks_example}). Since KeySwitch is the most expensive operation, the task and data distribution approach aims to optimize this particular operation.  For simplicity, KeySwitch for $dnum=L+1$ ($K=1$) \cite{KimPP22,HK20} is utilized throughout the paper.

\subsection{Monolithic vs Chiplet packaging}\label{sec:bg_packaging}
In the context of large Integrated Circuits, authors in \cite{chiplet_vs_mono,chiplet_natalie,chiplet_size_MaWWCH22} discuss the advantages of chiplet-based designs over monolithic designs. The problem with monolithic designs stems from the fact that to keep up with the increasing demand for high performance and functionality, chips need to be scaled up, and advanced technology nodes must be utilized. Manufacturing such big chips reduces the wafer yield as more surface area is exposed to defects per chip and increases the development cost. Such huge designs take a long time-to-market, and it is not straightforward to test and verify them. Factors such as size limitation and sub-optimal die performance due to overload contribute to a shift to SiP ~\cite{chiplet_vs_mono}. 

In SiP, multiple heterogeneous smaller chiplets can be manufactured separately and later integrated together using various packaging techniques. This promotes chiplet-reuse, lowering the development costs. The chiplet-packaging techniques can be broadly classified into three main categories: 2D, 2.5D, and 3D~\cite{chiplet_vs_mono,chiplet_size_MaWWCH22}. In 2D packaging, different dies are mounted on a substrate, known as a multi-chip module. Due to substrate limitations, this results in slow die-to-die communication and high power consumption.

To address these limitations, silicon interposers are used, and this technique is known as 2.5D integration~\cite{C2Ccomm1,occamy}. In this approach, an interposer is placed between the die and the substrate, enabling die-to-die connections on the interposer itself. The use of an interposer significantly enhances interconnectivity, leading to improved performance. 
Taking the integration capabilities a step further, 3D packaging involves stacking different dies on top of each other, akin to a skyscraper. In 3D packaging, the dies are interconnected using through-silicon vias (TSVs). 3DIC is gaining significant popularity and serves as the foundation for advancements like High Bandwidth Memory (HBM/HBM2/HBM3)~\cite{HBM3NoC,HBM3NoC1,3DICFFT,3D-DFT}. This approach significantly reduces the critical path, resulting in higher performance, lower power consumption, and increased bandwidth. The slowdown of Moore's law finds hope in 2.5D and 3D IC.

\subsection{NTT Design Techniques}

The multiplication of large degree polynomials is one of the major performance bottlenecks for FHE implementations. The number theoretic transform enables fast polynomial multiplications by reducing the complexity of polynomial multiplication to $\mathcal{O}(N \cdot \log N)$, and it is extensively employed for implementing FHE schemes. The NTT is defined as the Discrete Fourier Transform over $\mathbb{Z}_{q_i}$. As shown in \autoref{alg:cooley_tukey}, an $N$-point NTT operation transforms a polynomial ${a}$ of degree $N-1$ degree polynomial into another $N-1$ degree polynomial ${\tilde{a}}$. The NTT uses the powers of $N$-th root of unity $\omega$ (also referred to as twiddle factors) which satisfies $\omega^N \equiv 1 \pmod{q}$ and $\omega^i \neq 1 \pmod{q}$ $\forall i < N$, where $q \equiv 1 \pmod N$. Here, $q$ represents an RNS base, $q_i$. 
Similarly, inverse NTT (INTT) follows the same method with the modular inverse of $\omega$, and the resulting coefficients should be scaled by $1/N$. Prior FHE hardware acceleration works mainly utilized three techniques for implementing an NTT. The major difference between works is how they instantiate the butterfly unit highlighted in \autoref{alg:cooley_tukey}.

\begin{algorithm}[t]
    \renewcommand{\algorithmicrequire}{\textbf{In:}}
\renewcommand{\algorithmicensure}{\textbf{Out:}}
    
    \caption{The Cooley-Tukey NTT Algorithm \cite{DBLP:conf/ima/Scott17} }\label{alg:cooley_tukey}
    \begin{flushleft}
    \textbf{In:} A polynomial ${x}$ with coefficients $\{x_0, \cdots , x_{N-1}\}$ where $x_i \in \mathbb{Z}_{q}$, $N , q$
    
    \textbf{In:} Table of $2N^{th}$ roots of unity $\boldsymbol{g}$, in bit reversed order 
    
    \textbf{Out:} $\hat{{x}} \gets \texttt{NTT}({x})$, $\hat{{x}}_i \in \mathbb{Z}_{q}$, in bit-reversed order 
    \end{flushleft}
    \vspace{-10pt}
    \begin{algorithmic}[1]
    
    \STATE $t,m \gets (N/2), 1$\;
    
    \WHILE{($ m < N $)} 
        \STATE $k \gets 0$\;
        
        \FOR{($i=0$; $i<m$; $i=i+1$)}
    
            \FOR{($j=k$; $j<(k+l)$; $j=j+1$)}  
                \STATE $V \gets {x}[j + t] \times \boldsymbol{g}[m + i] \pmod q $\hfill \textcolor{gray}{$\triangleright$ Butterfly operation starts.}\\ 
                \STATE ${x}[j + t] \gets {x}[j] - V \pmod q $ 
                \STATE ${x}[j] \gets {x}[j] + V \pmod q $ \hfill \textcolor{gray}{$\triangleright$ Butterfly operation ends.}
                \ENDFOR
            \STATE $k \gets k+2t$\;
        \ENDFOR
        \STATE $t,m \gets t/2,2m$\;
    \ENDWHILE
    
    \RETURN ${x}$
    \end{algorithmic}
    \end{algorithm}

\begin{figure}[t]
    \centering
    \includegraphics[width=0.8\linewidth]{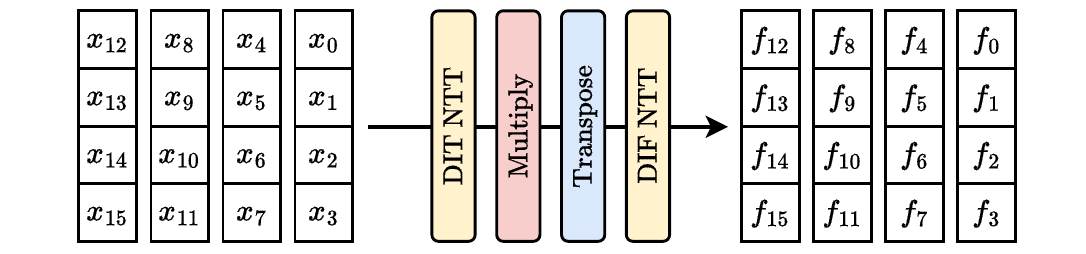}
    \caption{Hierarchical (4-step) NTT datapath for $N=16$. DIT stands for Decimation in Time, and DIF stands for Decimation in Frequency.}
    \label{fig:four_step}
\end{figure}

\begin{itemize}
    \item \textbf{Iterative.}  In this type of implementation, the prior works \cite{medha,ntt_intt} instantiate multiple butterfly units to process the inner `for' loop (\autoref{alg:cooley_tukey}) simultaneously, hence faster. Thus, the runtime of the NTT transformation for $m$ butterfly units is $\sim\frac{N\cdot\log{N}}{2\cdot m}$. 
    \item \textbf{Pipelined.} This is a bandwidth-efficient implementation where the butterfly units operate in pipelines instead of parallel. It is also used by prior works~\cite{zhang2021pipezk,pipentt} for applications such as zero-knowledge proofs where the polynomial degree is huge. It can be considered a stage unrolled NTT, where the unrolling is done based on the outer `for' loop instead of the inner loop (in \autoref{alg:cooley_tukey}). While its latency does not decrease, its throughput is much higher and useful for applications requiring multiple polynomial NTT transformations.
    \item \textbf{Hierarchical (4-Step).}  This implementation technique adopted by \cite{feldmann_2021f1}, predominantly utilizes the first technique as shown in \autoref{fig:four_step}. It splits one polynomial into a matrix of dimensions $N_1\times N_2=N$. The NTT units transform $N_1$ coefficients using multiple butterfly units instantiated in parallel, and there are $N_2$ such sets to process all $N_1\times N_2$ coefficients. After this first operation, the resultant data is transposed and multiplied with twiddle factors. Finally, other NTT units transform $N_2$ coefficients using multiple butterfly units instantiated in parallel, and there are $N_1$ such sets to process all the coefficients.  To save area $N_1, N_2$ are chosen so that $N_1=N_2$. This way, the same NTT unit can be utilized for both transformations ($N_1\times N_2$, $N_2\times N_1$). In this case, the transpose operation becomes complex and expensive as $N$ increases.
\end{itemize}

\begin{figure}[t]
    \centering
    \includegraphics[width=0.7\columnwidth,trim={0.5cm 0.2cm 2.8cm 0.4cm},clip]{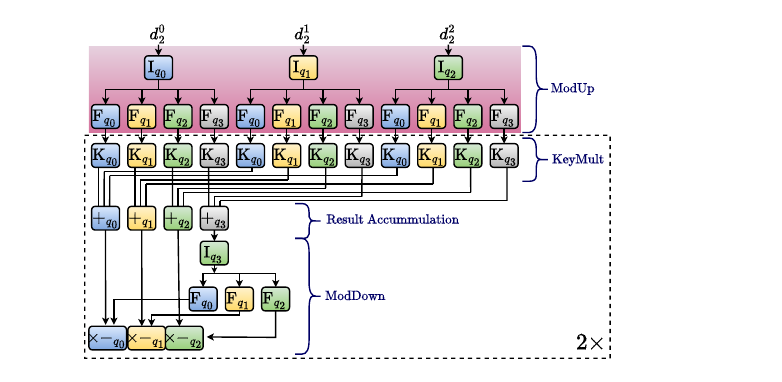}
    \caption{KeySwitch operation for $l=2$, where I, F, and K represent INTT, NTT, and key multiplication operations using MAS, respectively.}
    \label{fig:ks_example}
\end{figure}
\section{FHE-tailored Multi-Chiplet Design}\label{sec:mc_des}

A widely adopted approach for realising a chiplet-based architecture is to distribute the components of one large monolithic design across multiple chiplets connected in a mesh~\cite{mesh2,mesh1,mesh}. While such a disintegration approach has found utilities in applications like Machine Learning~\cite{ml_chiplet}, they fall short in leveraging the inherent algorithmic intricacies of FHE, thereby hindering efficient workload distribution among chiplets. 
In the following,  we investigate the trade-offs of various chiplet design possibilities for FHE and consolidate on an optimal solution. For this, let us consider the most performance-heavy macro routine- KeySwitch. Its data flow is illustrated in \autoref{fig:ks_example} for depth $l=2$.

\begin{figure}[t]
    \centering
    \includegraphics[width=\columnwidth,trim={1.4cm 0.8cm 0.4cm 0},clip]{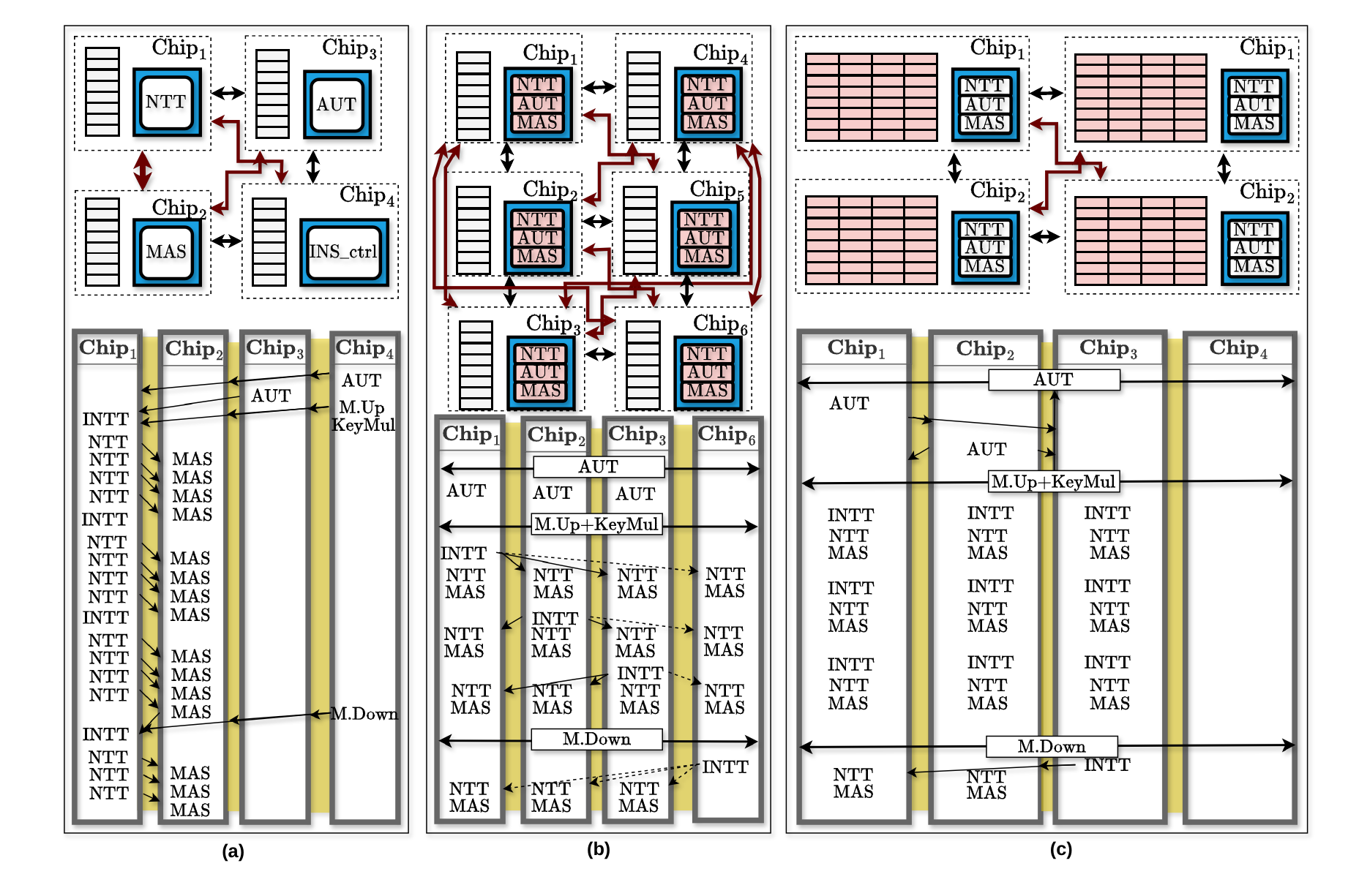}
    \caption{The diagram depicts the different techniques, data, and task distribution for automorphism followed by KeySwitch. $l=2,1$ for (a), (c) and $4$ for (b)}
    \label{fig:pior_approaches}
\end{figure}

\vspace{0.5em}\noindent \textbf{(a)} The naive approach for chiplet decomposition would be to closely follow the data flow of \autoref{fig:ks_example} and allocate one chiplet per square-box in the figure (I=INTT, F=NTT, K=Key Multiplication). This approach, as shown in \autoref{fig:pior_approaches} (a), will lead to $(i)$ uneven allocation of chiplet resources; for example, the chiplets for NTT or INTT will be much larger than those for MAS, $(ii)$ a massive increase in C2C communication overhead due to the continuous data exchange between the components, and $(iii)$ increase in required on-chip memory for data duplication. \autoref{fig:pior_approaches} (a) further provides a schedule for the KeySwitch operation depicted in \autoref{fig:ks_example}. In addition to illustrating the complexity of chiplet-to-chiplet communication and task dependencies, it demonstrates how most chiplets remain idle during computation, further emphasizing the imbalance in task distribution.
    
For a KeySwitch operation, the communication overhead is $(l+1)\cdot(l+2)$ polynomials for ModUp and $2\cdot(l+2)$ polynomials for ModDown. The time required for the operation is equivalent to the time taken to process $(l+1)$ INTT and $(l+1)\cdot (l+2)$ NTT operations for ModUp+KeyMul and $1$ INTT and $(l+1)$ NTT for ModDown of each ciphertext component. This is tabulated in \autoref{tab:tech_compare} and assumes that transfer can be done fast enough (monolithic) to compute MAS operations in parallel. If the $2\times$ slow communication cost is added, considering chiplet disintegration, the communication would take more time and surpass the computation overhead by $2\times$, thus becoming the major bottleneck. Hence, this approach significantly inhibits the performance of the design. A deeper disintegration would imply breaking the individual operation into several chiplets, for example, computing one NTT using several chiplets. Although this approach can lead to smaller chiplets, it will result in additional C2C communication overhead.

\vspace{0.5em}\noindent \textbf{(b)}  The second approach, illustrated in \autoref{fig:pior_approaches} (b), involves assigning each RNS limb computation, as depicted in \autoref{fig:ks_example}, to a single chiplet encompassing NTT, AUT, and MAS components. Computations with respect to one modulus are performed within the same chiplet. This approach also faces slow and more complex C2C communication overhead due to data dependencies between the processing of RNS limbs in the KeySwitch routine. Furthermore, as the depth of the FHE application decreases over time, reducing the number of RNS limbs ($l$), their respective chiplets become idle. 

\autoref{fig:pior_approaches} (b) illustrates the complexity of C2C communication with six chiplets, showcasing the all-to-all dependencies between the chiplets. This figure highlights that before the chiplets can begin evaluating the NTT, they must wait for the INTT operation to complete, which leads to certain chiplets sitting idle during this time. This idle time reduces the overall efficiency of the design and contributes to increased communication overhead due to the frequent data exchanges required between the chiplets. If we assume that the result can be immediately sent to all the PU (monolithic), the runtime can be quantified as $l+1$ INTT and $2\cdot(l+1)$ NTT(+MAS) for ModUP+KeyMul, and 1 INTT and 2 NTT(+MAS) for ModDown of each ciphertext component, as tabulated in \autoref{tab:tech_compare}. The total data in communication for ModUp is $(l+1)\cdot(l+2)$ polynomials, as all $l+1$ INTT resultant polynomials are broadcasted to $l+2$ chiplets. Similarly, for ModDown, the communication cost is $2\cdot(l+1)$ polynomials.  Hence, in a chiplet disintegration scenario where communication between chiplets is $2\times$ slower, it becomes the main bottleneck. 

\vspace{0.5em}\noindent \textbf{(c)}  A refined third approach, depicted in \autoref{fig:pior_approaches} (c), involves enabling each chiplet to support multiple RNS limbs to reduce C2C communication overhead and utilizing data duplication. For instance, the first chiplet can initiate NTT$_{q_1}$ after executing INTT$_{q_0}$ without sending it to the second chiplet, as depicted in \autoref{fig:ks_example}. This strategy minimizes communication overhead during ModUp and increases intermediate storage requirements by storing copies of input RNS limbs $d_2^0$, $d_2^1$, $d_2^2$ in \autoref{fig:ks_example} on each chiplet. INTT$_{q_0}$, INTT$_{q_1}$, NTT$_{q_0}$, etc., are computed within the same chiplet. As shown in \autoref{fig:pior_approaches} (c), this technique does not completely eliminate the complexity of C2C interconnects, as some inter-chiplet communication remains necessary to ensure duplication. Additionally, this approach requires significantly more on-chip memory since data must be duplicated across chiplets to enable local processing of multiple tasks, further increasing resource demands. This technique reduces wait time during ModUp but requires an all-to-all broadcast at the end of the computation. Hence, requiring transfer of $(l+1)\cdot(l+4)$ polynomials, including broadcast required for ModDown.\\\\
Thus, all the available approaches offer certain trade-offs and highlight communication as the major bottleneck, and our aim is to determine the best and most practical solution. With this aim, we develop a chiplet-based design approach for \papertitle.

\begin{table}[t]
    \centering
    \begin{tabular}{c|c|c|c|c|c}
    \hline
       & \textbf{Polynomials in} & \multicolumn{3}{c|}{\textbf{Computation}} & \multirow{2}{*}{\textbf{$\#$ Chiplets}} \\ \cline{3-5}
       & \textbf{Communication} & INTT & NTT & MAS \\
       \hline\hline
       Tech. (a)  & $(l+3)\cdot(l+2)$ & $l+3$ & $(l+1)\cdot(l+4)$ & || & 4 \\ \hline
       Tech. (b)  & $(l+1)\cdot(l+4)$ & $2$ & $l+4$ & $l+4$ & $L+2$\\\hline
       Tech. (c)  & $(l+1)\cdot(l+4)$ & $l+3$ & $l+4$ & $l+4$ & $L+2$\\\hline
       Our Tech.$^\dag$  & $4\cdot(l+3)$ & $\frac{l+1}{4}+2$ & $\frac{(l+1)\cdot(l+4)}{4}$ & || & 4 \\\hline
    \end{tabular}
    \caption{Comparison of the naive techniques and our technique for the KeySwitch operation, including ModUp, Key Multiplication, and ModDown. The operation count is the maximum reported by any chiplet, which will determine the overall throughput of the design. || Implies Operation is done in parallel with other computations.}
    \label{tab:tech_compare}
    \\$^\dag$ This technique is discussed in \autoref{sec:data_and_task_dist} for four chiplets ($r=4$).
\end{table}

\subsection{\papertitle 2.5D Architecture} \label{sec:2.5D_Arch}

In a chiplet-oriented design process, there are two critical choices: the size of chiplets and the number of chiplets. The manufacturing cost is reduced, and yield increases when the chiplets are small in area. However, having many small chiplets reduces performance as the complexity of slow C2C communication increases. In this work, we will develop a chiplet design strategy and integration topology, considering the data flow of FHE. This approach provides a balance between yield, manufacturing cost, and C2C communication overhead. 

\begin{figure}[t]
    \centering
    \includegraphics[width=0.8\columnwidth]{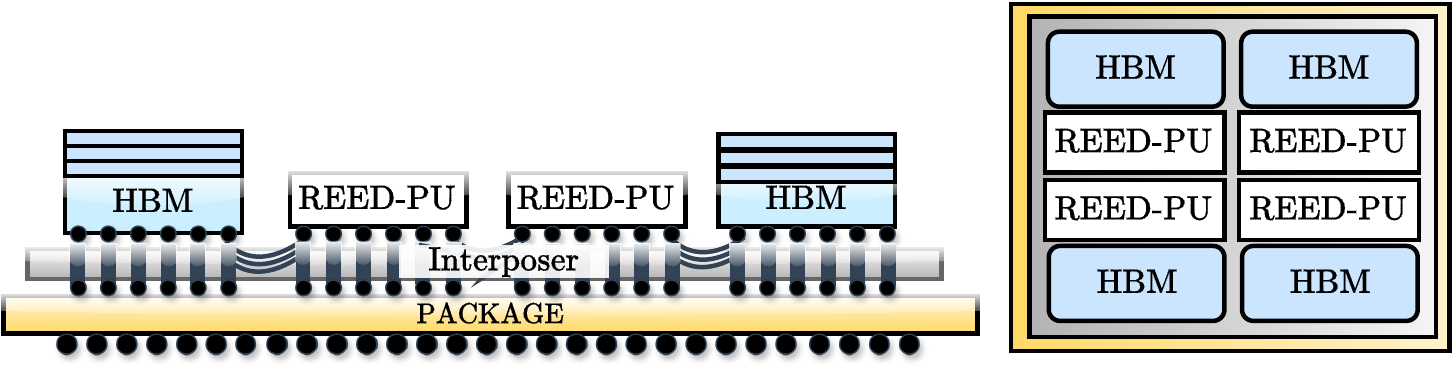}
    \caption{Side and top view of proposed four chiplet-based \papertitle 2.5D.}
    \label{fig:2_4dreed}
\end{figure}

As an example, \autoref{fig:2_4dreed} shows a four chiplet-based \papertitle 2.5D architecture, where the chiplets are connected in a ring formation and have exclusive read/write access to HBM in its proximity. Later, we will show that this architecture scales well with an increasing number of chiplets (\autoref{sec:higherchiplet}). 
To overcome C2C communication overhead and memory storage issues, we propose an RNS polynomial (limb)-oriented task and data distribution strategy, which is built on top of the third approach. Specifically, chiplets are assigned certain RNS limbs and all tasks related to these limbs without requiring data duplication (detailed in \autoref{sec:data_and_task_dist}).
The proposed ring formation (\autoref{sec:non_blocking_comm}) allows us to increase the number of chiplets at the cost of only a linear increase in the number of interconnects. Hence, we can scale it to eight or sixteen chiplets as well. This ring formation for connecting the FHE chiplets is specifically tailored to the data-flow of performance-critical FHE workloads. With this formation, not all dies need to communicate with every other die simultaneously, which is crucial for minimizing C2C communication requirements. 

Furthermore, our communication protocol ensures (\autoref{sec:non_blocking_comm}) that no HBM-to-HBM communication is required. Hence, HBMs are positioned on the outer side. We also avoid sharing one HBM among multiple chiplets, ensuring that each HBM is located only in proximity to the one chiplet it serves. Notably, our placement strategy aligns well with~\cite{occamy,manticore}, where authors design chiplet-based general-purpose processors with an actual tapeout, demonstrating practical viability. Finally, we also ensure a homogeneous design where all chiplets are identical, simplifying post-silicon realization.

\vspace{1em}


\noindent \textbf{Disintegration Granularity:} Chiplet systems face a trade-off between development cost and performance degradation, depending on the disintegration granularity. Existing works on chiplet-based architectures, such as \cite{3D-DFT,3DICFFT,chiplet_natalie,manticore,chiplet_size_MaWWCH22,chiplet_size_GPG}, show that disintegration improves yield, but it introduces challenges such as floorplanning and post-silicon testing overhead. Hence, the question:

\begin{center}
    \textit{How much disintegration is too much disintegration?}
\end{center} 

Considering a maximum die area of 800mm$^2$, dividing it into four chiplets offers an ${\approx}{80}\%$ yield, while eight chiplets provide a yield of ${\approx}{90}\%$. The yield numbers are obtained from \cite{chiplet_size_MaWWCH22}. 

While the eight-chiplet option shows promise for achieving high yield, it faces the challenge of underutilization over time as the number of RNS limbs decreases after rescaling. Specifically, when $l$ becomes smaller than 8, certain \papertitle chiplets remain idle (Detailed in \autoref{sec:data_dist} and \autoref{sec:higherchiplet}). On the contrary, the instantiation of four chiplets strikes an optimal balance between manufacturing cost and utilization. We want to remark that the number of \papertitle chiplets is flexible and can be changed as per user requirements, depending on the technology and computation constraints.

\section{Architecture Design of One Chiplet}\label{sec:design_methodology}

The need for scalability and high throughput drives our design methodology. We introduce the \papertitle design configuration- ($N_1,N_2$) for polynomial degree $N$, where $N_1\cdot N_2 = N$. For the clock frequency $f$, this configuration provides a throughput of $\frac{f}{N_1}$ operations per second and can process $N_2$ coefficients in parallel. A configuration-flexible design approach will help obtain computation-communication parallelism within every chiplet by ensuring that the memory read/write throughput is the same as the computational throughput. Now, let us explore how we design the ingredients of \papertitle Processing Unit (PU) to ensure flexibility.

\subsection{The Hybrid NTT (Frankenstein's Approach)}\label{sec:ntt}

\begin{figure}[t]
    \centering
    \includegraphics[width=0.85\columnwidth,trim={0.25cm 0.05cm 0.15cm 0.05cm},clip]{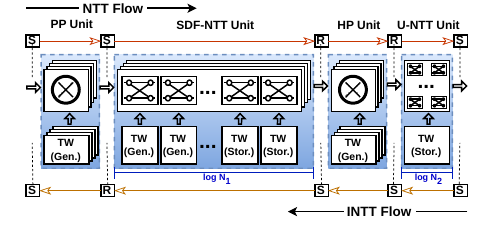}
    \caption{Novel routing-friendly Hybrid NTT/INTT design flow for $N=N_1 \times N_2$.}
    \label{fig:ntt}
\end{figure} 

The NTT/INTT unit plays a vital role in converting polynomials from slot to coefficient representation and vice versa. It is the most computationally expensive micro building-block and occupies over 50\% architectural area. Therefore, designing an efficient NTT/INTT unit is crucial as it directly impacts the overall throughput and area consumption of \papertitle. 

\vspace{1em}

\noindent \textbf{Prior works:} There are various approaches in the literature to implement NTT in hardware for large-degree polynomials, such as  iterative~\cite{medha,ntt_intt}, pipelined~\cite{zhang2021pipezk,pipentt} and hierarchical~\cite{feldmann_2021f1}, thoroughly discussed in \autoref{sec:background}. The implementation complexity of the plain iterative approach increases significantly with the number of processing elements. The pipelined approach (also known as single-path delay feedback (SDF)) provides a bandwidth-efficient solution but a diminished performance. The hierarchical approach (also referred to as four-step NTT), utilized in~\cite{feldmann_2021f1}, treats a polynomial of size $N$ as an $N=N_1 \times N_2$ matrix and divides a large NTT into smaller parts. It involves performing $N_1$-point NTTs on the $N_2$ columns of the matrix, then multiplying each coefficient by $\omega^{i\cdot j}$ (where $i$ and $j$ are matrix row and column indices), transposing the matrix, and finally performing $N_2$-point NTTs on the $N_1$ columns. 

Transposing a matrix of size $N_1\times N_2$ requires $N_1$ separate memories and large data re-ordering units. Hence, in \cite{feldmann_2021f1}, the transpose unit consumes 14\% of the area per compute cluster. Moreover, it also requires additional $N_2$ cycles for writing data to the transpose memory and $N_1$ cycles for reading it. Therefore, although the hierarchical approach simplifies the NTT implementation, we observe that it has the following limitations: $(i)$ the costly transpose operation, $(ii)$ fixed $N_1$ and $N_2$ such that $N_1=N_2$~\cite{feldmann_2021f1,basalisc}, and $(iii)$  the reliance on scratchpad in some works leads to large memory fan-in and fan-out, causing routing inefficiencies.

\begin{figure}[t]
    \centering
    \includegraphics[width=0.82\columnwidth,trim={0.7cm 0 32.8cm 0},clip]{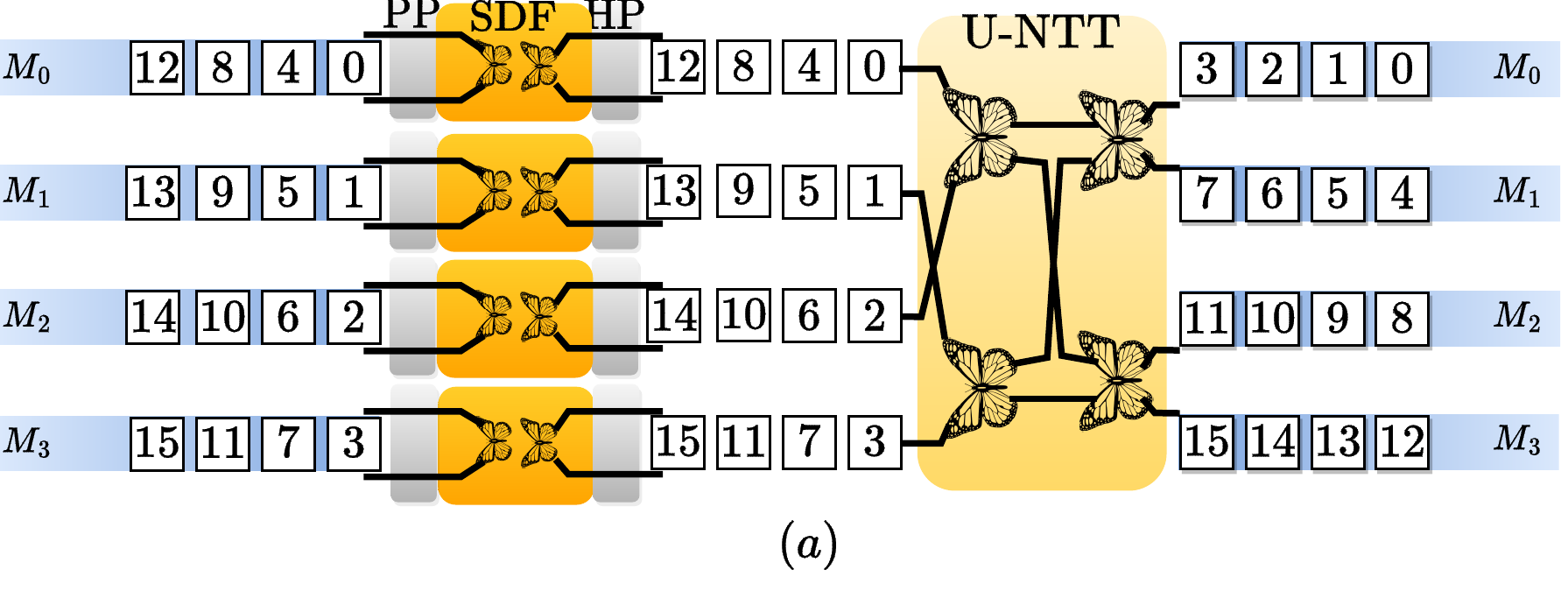}
    \includegraphics[width=0.82\columnwidth,trim={32.7cm 0 0.8cm 0},clip]{images/ntt_intt_mem_flow_2.pdf}
    \caption{The proposed novel Hybrid NTT/INTT design flow with Memory access for (a) NTT and (b) INTT with $N_1=4$, $N_2=4$, and $N=16$. The butterflies represent the Gentleman-Sande butterfly~\cite{ntt_intt} operation employed in our design. }
    \label{fig:ntt_hybrid}
\end{figure}

 \begin{algorithm}[t]
    \caption{Hybrid NTT with NWC}
    \label{alg:Hybrid_ntt}
        \begin{flushleft}
            \textbf{In:} ${a}$ (a matrix of size $N_1 \times N_2$ in row-major order)
        
        \textbf{In:} $\omega$ ($N$-th root of unity), $\psi$ ($2N$-th root of unity)
        
        \textbf{Out:} ${\tilde{a}} = \texttt{NTT}({a})$ (a matrix of size $N_1 \times N_2$ in column-major order)
        \end{flushleft}
        \vspace{-1em} 
    \begin{algorithmic}[1]
        \FOR{($i=0; i<N_1; i=i+1$)}
            \FOR{($j=0; j<N_2; j=j+1$)}
                \STATE ${a}[i][j] \gets {a}[i][j] \cdot \psi^{i\cdot N_2 + j} \pmod{q}$ \hfill \textcolor{gray}{$\triangleright$ PP:Pre-processing}
            \ENDFOR
        \ENDFOR
        
       \STATE  Apply $N_1$-pt NTT to the columns of ${a}$ \hfill \textcolor{gray}{$\triangleright$ using SDF-NTT }
        
        \FOR{($i=0; i<N_1; i=i+1$)}
            \FOR{($j=0; j<N_2; j=j+1$)}
                \STATE ${a}[i][j] \gets {a}[i][j] \cdot \omega^{i\cdot j} \pmod{q}$ \hfill \textcolor{gray}{$\triangleright$ HP:Hadamard prod.}
            \ENDFOR
        \ENDFOR
        
        \STATE Apply $N_2$-pt NTT to the rows of ${a}$ \hfill \textcolor{gray}{$\triangleright$ Unrolled(U)-NTT }  
        \RETURN ${a}$
    \end{algorithmic}
\end{algorithm}

\vspace{1em}

\noindent \textbf{Our Technique:}  We address these problems and propose a novel, scalable, and efficient NTT/INTT architecture. Our proposed design methodology-Hybrid NTT/INTT divides large NTTs into smaller parts \textit{that may differ in size}. To eliminate the transpose operation and reduce implementation complexity, we use a different design approach that amalgamates the two approaches mentioned above, thus the name \textit{Hybrid} NTT/INTT. A transpose is required in the hierarchical implementation approach to feed the data from unrolled NTT units back to the same unrolled NTT units. This is also the reason they need to have the same configuration ($N_1=N_2$). If, instead, we use pipelined NTT to replace one of the unrolled NTTs, no transpose will be required. This also results in design flexibility regarding $N_1$ and $N_2$ values.

 We utilize this idea to design an NTT that is routing-friendly, throughput-oriented, and does not necessitate costly transposition. To achieve this, we utilize parts of hierarchical, iterative, pipelined, and plain unrolled NTTs (Frankenstein's approach) and introduce a novel Hybrid NTT. It is fully pipelined, and its flow is shown in~\autoref{alg:Hybrid_ntt} and \autoref{fig:ntt}. The input polynomial for NTT operation is stored in $N_2$ memories of depth $N_1$, hence forming a matrix of size $N_1 \times N_2$ in row-major order. The NTT unit is fully pipelined and reads $N_2$ coefficients from memories in each cycle. We illustrated the memory layout and read the pattern of a polynomial during NTT operation for $N=4 \times 4$ in \autoref{fig:ntt_hybrid}. NTT unit first performs pre-processing (Step~3 of \autoref{alg:Hybrid_ntt}) using $N_2$ modular multipliers (PP unit), where each coefficient is multiplied with the corresponding twiddle factor.

The resulting $N_2$ coefficients represent one coefficient from each column of the input matrix, which should be processed via the $N_1$-pt NTT operation as shown in Step~6 of \autoref{alg:Hybrid_ntt}. To avoid in-between memory communication and reduce performance, we instantiate $N_2$ $N_1$-pt single-delay feedback (SDF)-based NTT units to perform $N_2$ NTT operations in parallel. Each SDF-based NTT architecture consumes and generates one coefficient per cycle after filling the pipeline. It utilizes $\log_2N_1$ cascaded butterfly units where each butterfly is coupled with a FIFO for data re-ordering~\cite{zhang2021pipezk}.

After the SDF-NTT unit, the resulting coefficients are multiplied with powers of $\omega$ as shown in Step~9 of \autoref{alg:Hybrid_ntt}. Similar to the PP unit, we use $N_2$ modular multipliers to perform this Hadamard product (HP unit). After this step, the hierarchical approach would require a transpose operation. To eliminate the transpose, we employ one fully unrolled $N_2$-pt NTT architecture (N-NTT unit) that takes $N_2$ coefficients as input per cycle and generates $N_2$ coefficients as output per cycle while performing a cost-free natural transpose operation. It does not incur an extra run-time cost as it is merged with the final step ($N_1$ $N_2$-pt NTTs).  Thus, the proposed Hybrid design reads $N_2$ coefficients per cycle and generates $N_2$ coefficients per cycle after filling the pipeline. It is scalable and enables different area/performance configurations by changing $N_1$ and $N_2$ values. This unit features bi-directional flow, and based on the configuration parameters, it provides a throughput of $\frac{f}{N_1}$. Overall, the properties and advantages of the proposed unit are as follows.

\begin{itemize}
    \item  \textbf{Transpose elimination:} The Hybrid NTT eliminates transpose by using two orthogonal NTT approaches, pipelined (SDF) approach for $N_1$-sized NTTs and unrolled (U-NTT) approach for $N_2$-sized NTTs. As shown in \autoref{fig:ntt_hybrid} (a), the output coefficients of SDF-NTT are processed directly by U-NTT, providing a seamless, natural transpose operation.
    \item \textbf{Bi-directional workflow:} The above method of transpose elimination also helps make our NTT unit bi-directional (for INTT), as illustrated in \autoref{fig:ntt_hybrid} (b). The additional routing complexity is balanced with efficient pipelining.
    \item \textbf{Low-level optimizations:} For modular multiplication and reduction unit, we adopted the word-level Montgomery~\cite{montgomery1985modular,mert2019design} modular reduction algorithm. and optimized it for our special prime form, $2^{w-1} + q_H \cdot 2^m + 1$, where $m=18$ is Montgomery reduction size, and $\lceil \log_2q_H \rceil=10$ is small. A total of $ (N_2+1)\log_2({N1})+\frac{N_2}{2}(\log_2{(\frac{N_2}{2})}+5)-7$ modular multipliers are utilized. 
    \item \textbf{On-the-fly twiddle generation:} We employ commonly used on-the-fly twiddle generation with a small constant memory that stores a few initial roots of the unity. This helps reduce the on-chip constant storage by up to 98.3\%.
\end{itemize}
 
One \papertitle-NTT requires $((N_2+1)\cdot \log_2(N_1)+(\frac{N_2}{2})\cdot \log_2(\frac{N_2}{2})+(5\cdot \frac{N_2}{2})-7)$ multipliers. The proposed transpose-less NTT also finds applications in other cryptographic schemes. For example, in Zero-Knowledge proofs~\cite{zkches}, where higher polynomial degrees make transposing more expensive, our NTT design offers a better area-time tradeoff.

\subsection{Multiply-Add-Subtract (MAS) and Automorphism (AUT)}
\begin{figure}[ht]
    \centering
    \includegraphics[width=0.55\columnwidth,trim={0 0.5cm 0 0.5cm},clip]{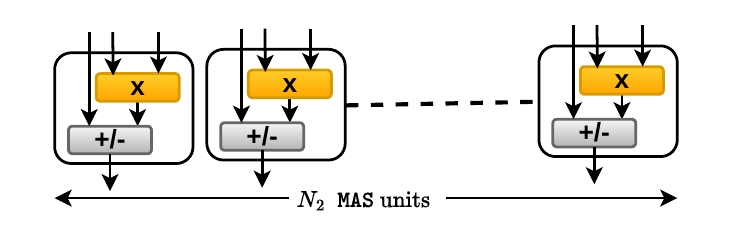}
    \caption{The set of triadic MAS units.}
    \label{fig:mac}
\end{figure}

MAS is elementarily designed as a \textit{triadic} unit for computing point-wise multiplication, addition, subtraction, or multiply-and-accumulate operations. It utilizes the modular multiplier proposed for the Hybrid NTT unit. A high-level overview of the MAS unit is shown in \autoref{fig:mac}. On the other hand, designing an efficient AUT unit is challenging~\cite{feldmann_2021f1,CraterLake}. It permutes ciphertexts using the Galois element ($gle$) to achieve rotation or conjugation. A polynomial is stored as a matrix $N_1\times N_2$ in $N_2$ memories.

\begin{algorithm}[t]
\caption{Automorphism} 
\label{alg:auto}

\begin{flushleft}
\textbf{In:} ${a}[N_1][N_2], gle $ 

 \textbf{Out:} $\hat{a} = \rho({a})$
 
\end{flushleft}
\vspace{-1em}
\begin{algorithmic}[1]
\STATE $index \gets gle$
\FOR{($l_0=0;l_0<N_1;l_0=l_0+1$)} 
    \STATE  $l_1 \gets index \pmod{\log(N_1)}$
    \STATE $start \gets index \gg \log(N_1)$
    \STATE $addr[j] \gets (start+j\cdot gle) \pmod{\log(N_2)}~ \forall ~j \in [0,N_2)$
   \STATE $ \hat{a}[l_1] \gets$ \texttt{shuffle\_tree\_}$N_2\times N_2(addr,{a}[l_0])$
   \STATE $index \gets index+gle$
\ENDFOR
\RETURN $\hat{a}[N_1][N_2]$
\end{algorithmic}
\end{algorithm}

\begin{figure}[t]
    \centering
    \includegraphics[width=0.5\columnwidth,trim={0 0.2cm 0 0}, clip]{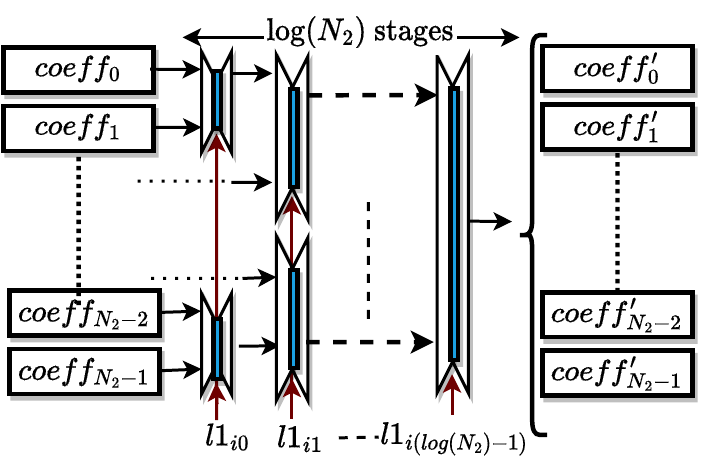}
    \caption{An example of a \texttt{shuffle\_tree\_$N_2 \times N_2$} workflow. Every stage has sufficient registers to hold $N_2$  coefficients.}
    \label{fig:shuffle}
\end{figure}

 For AUT, we make a key observation that when we load $N_2$ coefficients from memory address $l_0$ across all $N_2$ memories, they are shuffled based on the desired rotation offset ($\rho_{\textit{rot}}$), and then written to address $l_1$ across all $N_2$ memories. Hence, even though the coefficient order is shuffled, they all go to the same address of $N_2$ distinct memories. We utilize this property to permute all $N_2$ coefficients in parallel. This out-of-place automorphism is presented in~\autoref{alg:auto}. The in-place permutation techniques proposed in previous works~\cite{feldmann_2021f1,CraterLake} increase routing complexity due to memory transposition requirements. We are still left with a quadratically complex and expensive shuffle $\mathcal{O}(N_2^2)$ among the coefficients. However, we analyzed that all the shuffles could be performed pairwise on the coefficient batches, as shown in \autoref{fig:shuffle}. After each stage, two batches of coefficients are merged to form a new batch. We replace the naive and expensive operation with a pipelined binary-tree-like shuffle. Its number of pipeline stages adjusts with $N_2$, making the pipelining scalable and efficient for higher configurations. Moreover, the unit can handle any arbitrary rotation.

\subsection{PRNG-Based Partial Key-Switching Key Generation}\label{sec:prnggen}
 A PRNG is deployed to generate half the key components on the fly. The idea here is that the $ksk_1=a$ component of the key is generated by expanding a public seed and is assumed to be in NTT form, meaning it does not undergo an NTT transformation. Therefore, instead of storing the complete polynomial, the cloud or server can store the seeds of the key and generate them on the fly when needed. Note that the polynomials are generated in RNS form and have no interdependency. In other words, for ciphertext at depth, $l$, the $(l+1)\cdot(l+2)$ seeds are expanded, and the remaining limbs are ignored.  
 
 Our implementation uses the Trivium unit for PRNG similar to~\cite{medha}, which takes a 64-bit seed as input and initializes over 18 clock cycles, performing 18 rounds of operations. After initialization, the Trivium core generates 64-bit pseudorandom data continuously, producing one word per cycle without stalling. Each PU is equipped with one Trivium unit. The keys are only used for Key Multiplication during KeySwitch and are directly fed to the MAS  unit, which simultaneously consumes the pseudorandom words generated by the Trivium cores. The on-the-fly generation of the $ksk_1$ substantially reduces the memory footprint for key-switching. By generating it during runtime, we cut the required bandwidth (HBM-to-chiplet) and storage space for the key-switching key by 50\%. This dynamic approach optimizes memory usage and increases the efficiency of the system for key-switching operations. With this, we conclude the design of micro procedures.

\subsection{Programmable Instruction-Set Architecture}\label{sec:ISA}

In this section, we discuss how to program the micro procedures (NTT, AUT, MAS) for high-level FHE routines. This is crucial for determining their placement in the architecture, which will be discussed in the subsequent section.

Prior works define a strict operation flow. This prevents adaptations to future changes in the FHE algorithms or routine flow. Noting this, we utilize an instruction-based architecture design technique~\cite{roy_hpca}, wherein a relatively small instruction controller programs the \papertitle-PU, manages the multiplexers, and collects `done' signals from these units. The instruction controller includes a small BRAM for storing instructions. These instructions are provided at the start of the computation, along with the input data. Once the user sends an execute command, the instruction controller manages the execution of all stored instructions. This setup gives users the flexibility to specify any desired set of instructions, such as only performing NTT processing, and allows customization of parameters. Two types of instructions handled by the instruction controller are \textit{micro} and \textit{macro}. Micro-instructions are low-level arithmetic procedures like NTT, INTT, point-wise modular addition, subtraction, multiplication, multiplication-and-accumulation, and automorphism. They are used to compose microcodes for realizing macro-instructions for homomorphic addition/multiplication, KeySwitch, rotation, and moddown. A bootstrapping is performed using these macro instructions.


We initiate the PU design, as shown in \autoref{fig:rpau}, and uncover two important design decisions. The first is regarding the placement of NTT/INTT and MAS/AUT units. The second deals with the problems associated with large on-chip memories utilised in prior works~\cite{ARK,SHARP} for storing keys.

\subsection{\papertitle Processing Unit (PU)}\label{sec:packed_reed}

\begin{figure}[t]
    \centering
    \includegraphics[width=0.9\columnwidth,trim={1cm 0 1cm 0},clip]{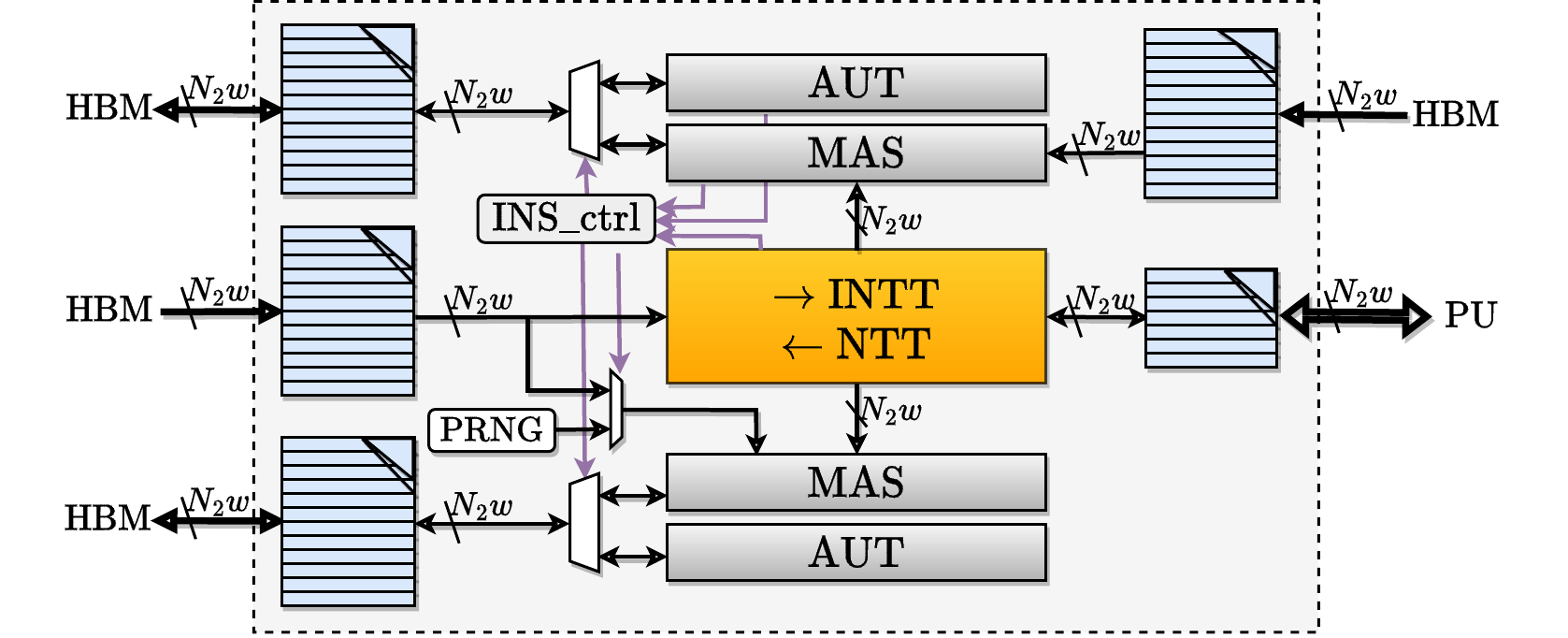}
    \caption{The \papertitle-PU design. Every data communication (memory to building blocks and off-chip to on-chip) here has a bandwidth of $N_2w$ bits/clock-cycle.}
    \label{fig:rpau}
\end{figure}

We initiate the PU design, as shown in \autoref{fig:rpau}. A PRNG is deployed to generate half the key components on the fly~\cite{medha}.  We uncover two important design decisions. The first is regarding the placement of NTT/INTT and MAS/AUT units. The second deals with the problems associated with large on-chip memories utilised in prior works~\cite{ARK,SHARP} for storing keys.

\vspace{0.5em}

\noindent \ballnumber{1} The polynomial processed by NTT unit is multiplied with two polynomials of KeySwitch keys and accumulated. Hence, we instantiate a pair of MAS/AUT units capable of simultaneously processing both key components.  Thus, the design has the ability to run NTT and MAS units concurrently  (shown in \autoref{fig:timeline}), which improves the KeySwitch performance by 66.7\%, as explained in \autoref{sec:th_ks}.  Moreover, since the AUT and MAS units are relatively cheaper, this design decision also does not add significant area overhead, as presented later in \autoref{tab:area}. 

\vspace{0.5em}

 \noindent \ballnumber{2} 
 In hardware accelerators, on-chip memory causes significant area overhead. Chiplets with large on-chip memory are not power, area, and manufacturing cost efficient~\cite{web_link7}. In the context of FHE, each KeySwitch key demands $~{\approx}{1}$MB or $91$MB storage for $dnum=L+1, \text{ and } L=30$ or $dnum=3, \text{ and } L=22$ respectively. Given the limited on-chip memory capacity of small chiplets, accommodating even a single KeySwitch key becomes rather challenging. Consequently, reliance on off-chip memory access becomes essential when a KeySwitch operation necessitates a different key. It is also not useful to store just one relinearization key required after ciphertext multiplication, as for a majority of applications, more rotations are required compared to multiplications~\cite{matmul_appl}. Therefore, in our architecture, we store all the keys in the large off-chip HBM. To reduce the overhead of off-chip memory access, we develop an efficient prefetch unit that streamlines data movements in parallel to computation, as described next. 

\subsection{Throughput Computation for KeySwitch}\label{sec:th_ks}

\begin{figure}[t]
    \centering
    \includegraphics[width=0.65\columnwidth, trim={ 1cm 0.25cm 0 0}, clip]{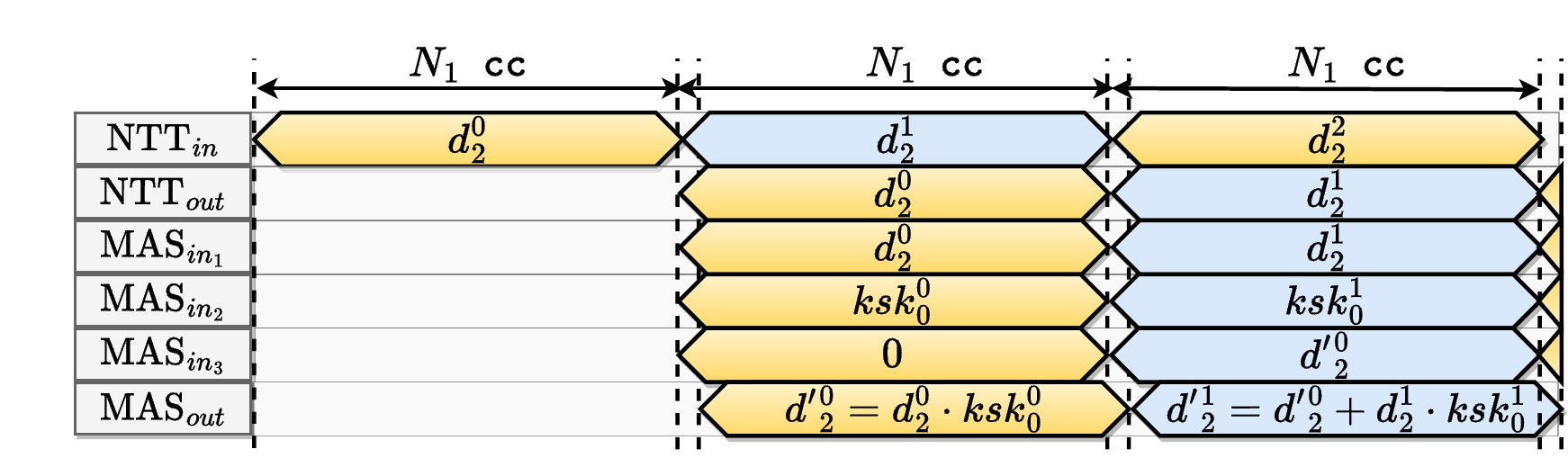}
    \caption{Timeline of parallel and pipelined operation flow.}
    \label{fig:timeline}
\end{figure}

The KeySwitch, detailed in \autoref{algo:relin_14}, is the most expensive operation among all the routines. For $dnum=L+1$, it transforms all $L+1$ residue polynomials from slot to coefficient representation (INTT), and then each of these is transformed to  $L+2$ NTTs, multiplied with two key components, and accumulated. This requires $L+1$ INTTs, $(L+1)(L+2)$ NTTs, and $2(L+1)(L+2)$ MAS, making the naive throughput of this operation: $\frac{f}{(L+1)(1+3(L+2))\cdot N_1}$. By utilizing \papertitle's parallel processing capability to perform all MAS operations concurrent to the NTT operations (shown in \autoref{fig:timeline}), we save $2(L+1)(L+2)$ clock cycles and increase the throughput to $\frac{f}{(L+1)(L+3)\cdot N_1}$, resulting in a 66.7\% improvement. 

\subsection{Streamlined Prefetch for On-Chip Storage}\label{sec:cach_mem} 
As mentioned in~\autoref{sec:ntt}, \papertitle's design methodology mitigates the need for scratchpad-like on-chip memory, allowing us to use memory units solely as prefetch units. 

Each memory unit in \autoref{fig:rpau} exhibits balanced fan-in and fan-out, and among the five memory units depicted, four are fed by off-chip memory. The small memory is responsible for storing and communicating the INTT result to the other PUs  (or Chiplets) (elaborated in \autoref{sec:non_blocking_comm}). Only two of the four memories communicating with off-chip memory need to write back the results, as illustrated by bi-directional arrows in \autoref{fig:rpau}. 

Summarily, three memories perform off-chip read/write communication. These memories are physically divided into two parts. When one is utilized for on-chip computation, the other performs off-chip prefetch (similar to ping-pong caching).

\section{Techniques for exploiting Comm-Comp Parallelism}\label{sec:data_and_task_dist}

The previous section discussed how a configuration-based design methodology enables off-chip and on-chip communication-computation parallelism. However, when distributing FHE workloads among multiple chiplets, we must consider the C2C communication overhead. This is important as state-of-the-art HBM3~\cite{hbmgen2} features a bandwidth of 1.2TB/s, while the state-of-the-art C2C interconnect UCIe~\cite{C2Ccomm1,C2Ccomm2} only offers a bandwidth of 0.63 TB/s. Consequently, a chiplet system optimized for HBM bandwidth could face bottlenecks due to slow C2C communication. 
The routine that necessitates most data exchange is the ModUp portion of KeySwitch, detailed in \autoref{alg:keyswitch}. It switches the modulus of each $L+1$ residue polynomial ($\texttt{INTT}$($d_2{_{q_i}})_{q_i}$  $~\forall~ i \in [0,L]$) to ($L+2$) residues polynomials   ($\texttt{NTT}(d_2{_{q_i}})_{q_j}~\forall~j \in [0,L+1]$). The ModUp operation is followed by key multiplication and ModDown. This is the flow for $dnum=L+1$, and a more detailed discussion of our proposed technique for $dnum<L+1$ is provided in \autoref{appendix:keyswitch}.

\subsection{Communication cost-analysis}
 In a multi-PU work~\cite{medha}, the authors briefly discuss limb-based decomposition and propose distributing computation across the RNS polynomials (limbs) by employing one PU per limb. While this approach enables highly parallel computations, as the multiplicative depth decreases, many PUs become idle, causing underutilization.
 
 In \cite{BTS_isca,ARK,SHARP}, the authors utilize both the limb-based and coefficient-based task distribution during KeySwitch.  They~\cite{ARK,SHARP} propose using limb-wise distribution for INTT and NTT steps and coefficient-wise distribution for the modulus switch (referred to as Base Conversion in the paper).  \cite{ARK} utilizes four PUs, and since thus base conversion is needed between the INTT and NTT steps, all-to-all broadcasts are done across PUs to switch from one distribution to another. In the subsequent subsection, we analyse how both techniques attain the same communication overhead.

Additionally, the all-to-all C2C broadcast between the chiplets is slow and increases by $\mathcal{O}(\mathtt{r}^2)$ with the number of chiplets $\mathtt{r}$. In the context of FHE, switching between limb- and coefficient-wise task distribution becomes expensive as it demands all-to-all C2C data movements. For example, \cite{ARK} proposes utilizing four PUs in a single monolithic chip. However, when extended to a chiplet setting, where each PU occupies a separate chiplet, the lack of an all-to-all broadcast capability makes it difficult to send data across all chiplets instantly. Using bi-directional C2C communication ability, the polynomial would reach all four chiplets via at least two serial C2C communication interfaces. The on-chip bandwidth used in the prior works is (20TB/s~\cite{ARK}, 36TB/s~\cite{SHARP}) is much less than the state-of-the-art C2C communication bandwidth (0.63TB/s~\cite{C2Ccomm1,C2Ccomm2}). As a result, chiplets would have to wait longer for data to arrive before computing, and this C2C communication overhead will significantly inhibit the performance.
Hence, there is a need to devise a schedule that can couple most of the communication with computation.

\subsection{Limb-wise vs Coefficient-wise distribution}\label{sec:lc_dist}

   \cite{ARK} utilizes four PUs and states that after the ModUp step, the number of polynomials that need to be transferred during limb-based only decomposition is $2\cdot dnum\cdot (L+K+1)$, while for coefficient-wise it is $(dnum+2)\cdot(L+K+1)$. They present this discussion in the context of a generalized KeySwitch technique for an arbitrary value of $dnum$, presented in \autoref{algo:relin_gen} (in \autoref{appendix:keyswitch}).
 
 This assumes the limb-wise distribution is done after multiplication with KeySwitch keys, and all the results are sent to every PU via an all-to-all broadcast. However, we remark that one PU does not need to send all the polynomials and instead only needs to send $\frac{2\cdot(dnum-1)\cdot(L+K+1)}{dnum}$ polynomials so that every PU holds the results for the supported bases. Therefore, the total cost becomes $2\cdot(dnum-1)\cdot(L+K+1)$, which is less than the cost of coefficient-wise distribution for $dnum=3$. Furthermore, we would like to note that polynomial distribution after KeySwitch is expensive as the data doubles in size after multiplication with the two key components.  Hence, this distribution should be done immediately after NTT computation, which further reduces the cost to only $(dnum-1)\cdot(L+K+1)$. This is much less compared to coefficient-wise distribution $(dnum+2)\cdot(L+K+1)$. Hence, we reassert that limb-based distribution will attain less communication overhead than coefficient-wise without any extra computation overhead.

\subsection{Data Distribution across Multiple Chiplets}\label{sec:data_dist}

The above analysis establishes that limb-based decomposition is the best task and data distribution technique across multiple chiplets. Within one chiplet, the computation is coefficient-wise distributed as $N_2$ coefficients are processed in parallel (discussed in~\autoref{sec:design_methodology}). However, as discussed in \autoref{sec:mc_des} (b) and (c), the limb-based distribution technique also has limitations. Therefore, we adapt it to offer long-term high performance.

This adaptation stems from the two key observations from the data flow of the KeySwitch operation in \autoref{fig:ks_example}. Firstly, the INTT results that need to be shared and duplicated are ephemeral, and thus, they do not require any long-term storage -- subsequent operations immediately consume them. To improve the efficiency, instead of duplicating the INTT data or sharing memory across chiplets, we leverage a \textit{ring-based C2C communication} as shown in \autoref{fig:ntt_comm}. The chiplets are connected in a ring formation where each chiplet processes one INTT result and then sends it to the next chiplet. The ring-based data movement minimizes the number of C2C interconnects and ensures that each chiplet operates on independent memory, simplifying placement and routing constraints.

The second observation is related to the underutilization of chiplets with reduced levels or depths in the ciphertext after homomorphic rescaling. If we distribute the limbs across the chiplets in an interleaved manner, then as the multiplication depth decreases, the number of limbs per ciphertext in each chiplet also uniformly decreases. To explain the benefits of the interleaved distribution, we will use the analogy of a card game.

\vspace{1.5em}
\noindent \textbf{The FHE card game:}  In this game, each player represents a chiplet, while the residue polynomials or limbs act as the cards. The cards dealt out are collected in a LIFO (last-in-first-out) manner~\footnote{It can also be FIFO (first-in-first-out) without any loss of generality.}, imitating the loss of multiplicative depth during computation. All players engage in the game (FHE routine computation) until they exhaust their cards. The start of a new FHE computation game mirrors Bootstrapping, which starts when only one player retains a single card. Until this point, players without cards must wait until the next game to participate. With these rules, the dealer (user or compiler~\cite{fhe_compiler}) has two choices: deal out all the cards to one player before moving to the next or do alternate distributions such that every other card goes to different players. Let us say there are $L=32$ cards and $\mathtt{r}=4$ players.  In the former case, the first player gets the cards drawn at instances $\{0,1,2,\cdots,7\}$, and in the latter case, at instances $\{0,4,8,\cdots,28\}$, and so on for the other players.

 In both scenarios, each player receives $8$ cards in total. In the first option, the last player exhausts their cards first, followed by the preceding player, and so on and so forth. Consequently, until the first player runs out of cards, others remain inactive. Conversely, with alternating distribution, each player loses one card in turn. Thus, at any point in the game, players either possess the same number of cards or one less, ensuring active involvement throughout.

The goal of FHE architecture design is to ensure the full utilization of chiplets in the long term and, thus, deliver high performance for the available computation resources. It translates to maximizing the player interaction in the FHE card game. Thus, the latter technique of interleaved alternate distribution offers maximum chiplet participation in the "card game" of computations.  Now, if there are too many players, then the number of players becoming idle will increase no matter which technique is used. The latter technique will only minimize the idle time. This is the problem with using more chiplets. Thus, next, we will discuss our final key technique for reducing communication overhead and then derive a good upper-bound on the number of chiplets.


\subsection{Efficient Non-Blocking C2C Communication}\label{sec:non_blocking_comm}

\begin{figure}[t]
    \centering
    \includegraphics[width=0.6\columnwidth,trim={0.25cm 0.15cm 0.25cm 0.1cm},clip]{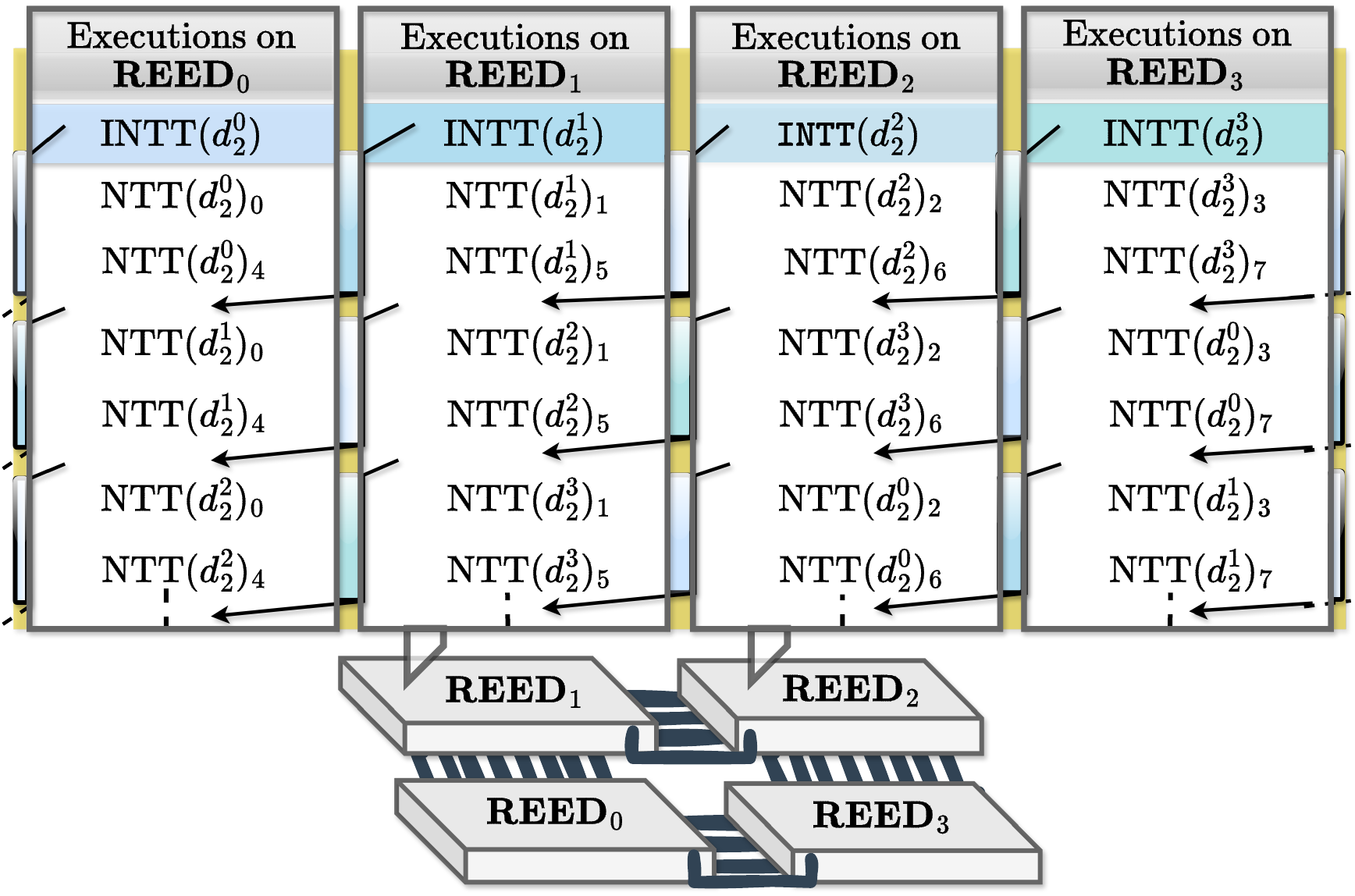}
    \caption{Non-blocking ring-based communication for four \papertitle chiplets when $l=6$. The blocks between executions represent the long communication window to make up for slow inter-chiplet (C2C) communication.}
    \label{fig:ntt_comm}
\end{figure}
\begin{algorithm}[t]
\caption{ModUp\_KeyMul} 
\label{alg:keyswitch}
\begin{flushleft}
\textbf{In:} $\boldsymbol{d_2} $  (the ciphertext component to be linearized)\\
\textbf{Out:} $\text{BUF} = ModUp\_KeyMul(\boldsymbol{d_2})$
\end{flushleft}
\vspace{-10pt}
\begin{algorithmic}[1]
\STATE Following tasks are executed by $\text{\papertitle}_i~\forall~i \in [0,r)$ 
\\ \textcolor{gray}{$\triangleright$ All $\text{\papertitle}_i$ operate \textbf{in parallel} as shown in \autoref{fig:ntt_comm}}
\FOR{($j=0;j<\frac{l+1}{r};j=j+1$)} 
    \STATE $ \mathtt{I^{rcv}_i} \gets \mathtt{INTT}(\boldsymbol{d}_2^{j\cdot r+i})$
    \\\textcolor{gray}{$\triangleright$ \textit{Initiate communication with} $\text{\papertitle}_{(i+1)\bmod{4}}$, $\text{\papertitle}_{(i-1)\bmod{4}}$}
    \FOR{($m=0;m<r;m=m+1$)} 
    \STATE $\mathtt{I^{proc}_i} \gets \mathtt{I^{rcv}_{i}}$
    \\ \textcolor{gray}{$\triangleright\triangleright$ \textit{Long Communication window opens now} $\triangleleft\triangleleft$}
    \\ \textcolor{gray}{$\triangleright$ \textit{Receive}  $\mathtt{I^{rcv}_{(i+1)\bmod{4}}}$ \textit{from} $\text{\papertitle}_{(i+1)\bmod{4}}$ \textit{and Send}  $\mathtt{I^{rcv}_i}$ \textit{to} $\text{\papertitle}_{(i-1)\bmod{4}}$}
    \FOR{($t=0;t<\frac{l+2}{r};t=t+1$)} 
    \STATE $\text{BUF}_{t*r+i}~+=~( \mathtt{NTT}(\mathtt{I^{proc}_i})\cdot \mathtt{KSK}^{j\cdot r +(i+m)\bmod{4}})_{q_{t*r+i}}$
    \ENDFOR
   \\\textcolor{gray}{$\triangleright\triangleright$ \textit{Ensure}  $\mathtt{I^{j\cdot r+(i+m+1)\bmod{4}}}$ \textit{has been received}}
   \\ \textcolor{gray}{$\triangleright\triangleright$ \textit{ Communication window closes} $\triangleleft\triangleleft$}
   \STATE $\mathtt{I^{rcv}_i} \gets \mathtt{I^{rcv}_{(i+1)\bmod{4}}}$
\ENDFOR
\ENDFOR
\RETURN $\text{BUF}$
\end{algorithmic}
\end{algorithm}

The proposed ring-based communication still faces overhead as the chiplets have to wait for every chiplet before them in the ring to send the INTT result. Hence, we propose a communication strategy, illustrated in \autoref{fig:ntt_comm}, to overcome this remaining problem.

The key idea is that the chiplets concurrently operate on different limbs instead of waiting for one limb and then processing it, as shown in \autoref{alg:keyswitch}. Each chiplet starts with the assigned limb, computes INTT, and then performs multiple NTTs on it. While performing NTT, it starts sending/receiving the INTT result. For example, the {REED$_0$} sends its INTT result to {REED$_{3}$} and receives the INTT result from {REED$_1$}. This is a uni-directional \textit{ring-based communication}. Since only one INTT result needs to be sent for  $\frac{l+2}{\mathtt{r}}$ parallel NTT computation, we have a larger C2C communication window compared to computation. Consequently, \emph{non-blocking communication} is achieved as data computation can proceed concurrently with relatively slower communication.

This technique necessitates just one read/write port per chiplet, in contrast to the requirement for ($\mathtt{r-1}$) ports in a star-like (i.e., all-to-all) C2C communication network.

\vspace{1.5em}
\noindent \textbf{The Adapted Data and Task Distribution Technique.}\vspace{0.5em}\\ C2C communication bandwidth plays a major role in the performance versus the number of chiplets trade-off. Let us assume the C2C communication bandwidth is k$\times$ slower than the HBM to Chiplet communication bandwidth. Each chiplet operates on $\frac{L+2}{\mathtt{r}}$ polynomials. The total computation time to process all $L+2$ polynomials for the KeySwitch should be close to $k$. Otherwise, we will not be able to decouple the communication from computation as discussed above.  This offers us a loose upper bound $ \mathtt{r}<\frac{L+2}{k} $. To ensure $u\times$ higher utilization, this bound must be made tighter. $u$ can take any value $\leq \frac{L+2}{k}$ ($u=\frac{L+2}{k}$ for monolithic chip). We take $u$ as 4. The adapted limb-based task/data distribution technique has the following properties: 
\begin{itemize}
    \item The number of chiplets is constrained by $ \mathtt{r}\leq \frac{L+2}{4k} $.
    \item An interleaved data/task distribution approach is utilized such that Chiplet$_i$ gets data and task corresponding to limbs ${\mathtt{r} j+i}$ ${\forall~0 \le j < \frac{L+2}{\mathtt{r}}}$, instead of sequential allotment (Chiplet$_i$ $\gets$ ${\mathtt{r} i+j}$).
    \item The technique outlined in \autoref{alg:keyswitch} is to be followed by all Chiplets to minimize data exchange overhead and costly C2C interconnects.
\end{itemize}

\subsection{ModDown/Rescaling Task flow}

\begin{figure}
    \centering
    \includegraphics[width=\linewidth,trim={1.1cm 0.4cm 1.7cm 0.3cm},clip]{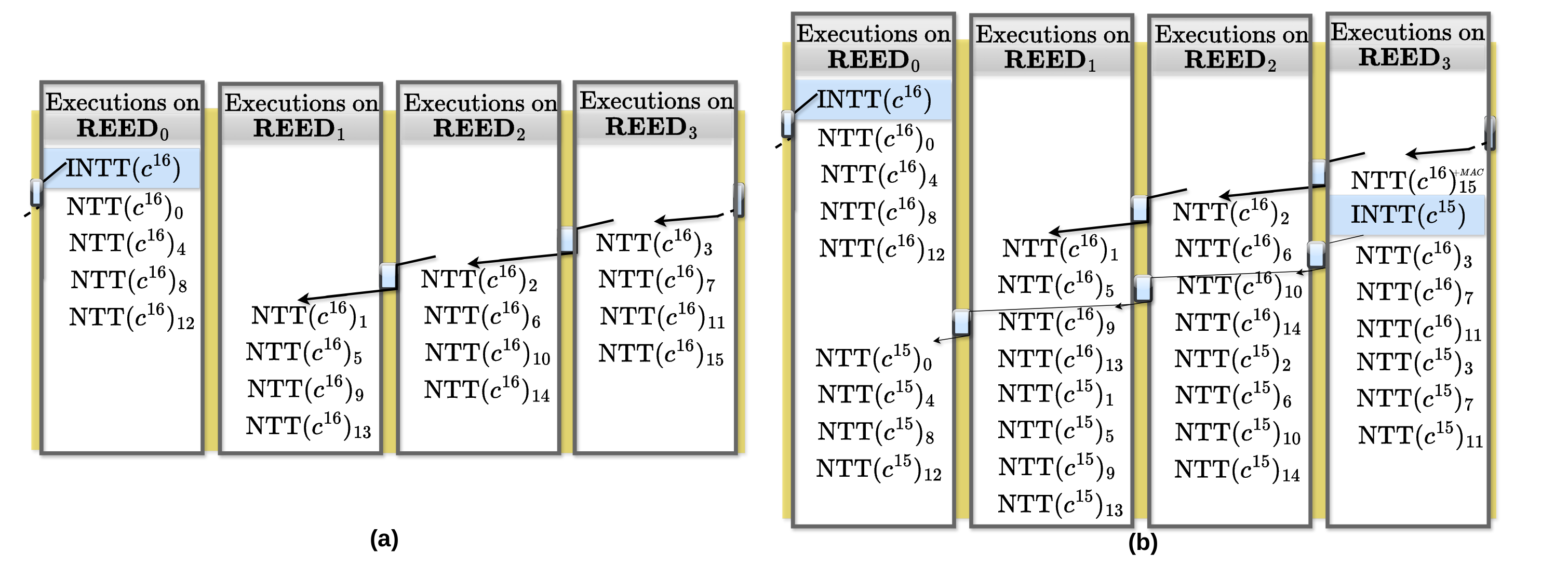}
    \caption{Ring-based communication for four \papertitle chiplets when $l=15\rightarrow16$ (after ModUp) for the (a) ModDown operation and (b) ModDown+Rescale operations.}
    \label{fig:moddown_comm}
\end{figure}

So far, we have discussed the data and task flow in the context of ModUp operation, which exhibits the highest complexity in terms of NTT operations. Next, we turn our attention to the ModDown/Rescaling operation.  As shown in \autoref{fig:ks_example}, the ModDown operation constitutes the final step of the KeySwitch procedure.  This operation involves a single INTT followed by $l+1$ NTT operations, resulting in significantly lower complexity compared to the ModUp operation. \autoref{fig:moddown_comm} (a) illustrates the communication flow for ModDown across multiple chiplets. Specifically, the chiplet that holds the polynomial corresponding to the modulus being dropped performs the INTT and sends the resulting data to the next chiplet. The data is then propagated through the remaining chiplets in a feed-forward ring topology, as depicted in the figure. Notably, the MAS operations are done simultaneously with the NTT operations.

While the overall computational complexity of the operation is $\frac{l+1}{r}$ NTTs per chiplet, there is an additional fixed communication overhead of $r-1$ polynomials due to inter-chiplet data transfer. A similar flow is observed in the Rescaling operation, which reduces ciphertext depth following noise growth. However, Rescaling is not strictly needed after the KeySwitch procedure and can be performed independently to mitigate noise after several rotations, accumulations, or plaintext multiplications. ModDown and Rescaling are applied to each ciphertext component, requiring two polynomial broadcasts. Note that the figure shows the processing of one component as the second component's communication will be done in parallel with the first component's computation. The INTT will be computed by the same chiplet and transferred, while the other chiplets compute the NTT for the previous component. Hence, no communication overhead will be incurred for the second component. Runtime of ModDown and Rescaling is included in the total runtime reported for the Relinearization operation (tabulated in \autoref{tab:perf}).

When Rescaling is required immediately after a ModDown operation, it leverages the optimized flow from the ModUp operation, as demonstrated in \autoref{fig:moddown_comm} (b). Both processes necessitate an INTT broadcast followed by NTT computation, where the chiplet executing the INTT performs one less NTT at best, effectively dropping the computation with respect to the modulus used for the INTT. Consequently, while the first set of NTT computations is underway, the second INTT result can be broadcast; thus, wait time overhead is incurred only once overall.

\section{Implementation Results}\label{sec:results}

\begin{figure}[t]
    \centering
    \includegraphics[width=0.55\columnwidth]{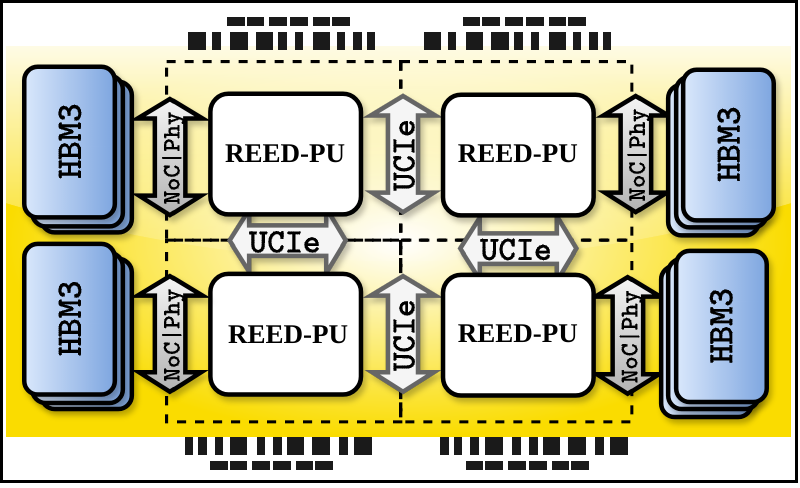}
    \caption{The complete architecture diagram of 4-chiplet \papertitle 2.5D for 1024$\times$64 configuration. The multiple small black blocks denote I/O interconnects along the edges.}
    \label{fig:reed_chiplet}
\end{figure}

\begin{table}[t]
\caption{Performance micro-benchmarks for 28nm and 7nm. }
\centering
\begin{tabular}{l|c|c|c}
    \hline
 \textbf{Micro-Benchmarks} $\downarrow$& \textbf{Level} & \multicolumn{2}{c}{ \textbf{Time (ms)} } \\ \cline{2-4}
  \textbf{Configuration} $\rightarrow$ & \textbf{l }    & \textbf{1024$\times$64 }  &   \textbf{512$\times$128}  \\ \hline\hline
     AUT/MAS (pt-ct)     & 30                & 0.005 &  0.003 \\ 
       MAS (ct-ct)       & 30                & 0.01 &  0.005 \\
        KeySwitch        & 30$\rightarrow$31 & 0.19  &  0.08 \\
        MULT \& Relin.   & 30$\rightarrow$29 & 0.22 &  0.11  \\
        Bootstrapping    & 1$\rightarrow$30$\rightarrow$15 & 14.2 &  7.1  \\\hline
\end{tabular}
\label{tab:perf}
\end{table}
Based on the precision-loss study done for $dnum=L+1$ (shown in \autoref{sec:prec_loss}), we choose the overall parameters for synthesizing and benchmarking our design as $N=2^{16}, L=30, K=1, L_{boot}=15, w=54$. Upon implementation (silicon realisation), only the parameters $N,w$ are fixed, and the other parameters (e.g., $dnum$) can be changed as per application requirements. For a higher value of $K$, the task distribution is discussed in \autoref{appendix:keyswitch}. The user can control this via the low-level instruction abstraction provided.

We synthesize the entire \papertitle 2.5D PU for configurations 1024$\times$64 and 512$\times$128 using TSMC 28nm and ASAP7~\cite{asap7} 7nm ASIC libraries with Cadence Genus 2019.11, and SRAMs are used for on-chip memories (with total storage capacity of 14 polynomials). The \papertitle-PU and all its building blocks are fully implemented using Verilog Hardware Definition Language. We simulated our design using Vivado 2022.2 for functionality testing.  The Instruction Controller is a part of the design and is connected to the master processor via the same master-slave interface as data in prior works~\cite{medha}.  Our primary objective is to achieve high performance while optimizing area and power consumption. To this end, we set our clock frequency target to 1.5 GHz, use High-vt cells (hvt) configuration for low leakage power, enable clock-gating, and set the optimization efforts to high. We set the input/output delays to 20\% of the target clock period and leverage incremental synthesis optimization features. Moreover, we take a step further by \emph{prototyping} the building blocks on Xilinx Alveo U250 to verify functional correctness, which has not been done by prior ASIC FHE accelerators.

As off-chip storage, we leverage the state-of-the-art HBM3 \cite{hbmgen2,jedec,HBM3NoC,hbm} memory, offering improved performance and reduced power. It is already deployed in commercial GPUs and CPUs~\cite{nvidia}. HBM3 with 8/12 stacks of 32Gb DRAMs has 32/48 GB storage capacity \cite{jedec,synopsys}. The ciphertexts provided by the client can be transferred to \papertitle using 32 lanes PCIe5 offering a bandwidth of 128 GB/s~\cite{pcie5}. In our work, we present results for HBM3 PHY and HBM3 NoC (Network On Chip), based on~\cite{hbmgen2,HBM3NoC,HBM3NoC1} with reported bandwidth of 1.2TB/s~\cite{hbmgen2}. For C2C communication, UCIe (Universal Chiplet Interconnect Express) advanced interconnect can offer a bandwidth of 0.63 TB/s \cite{C2Ccomm1,C2Ccomm2} for 2.5D integration. Thus, we can send or receive $64$, $54-bit$ coefficients per clock cycle. Overall we present the results using the following: (i) Verilog description of REED-PU, 
(ii) Instruction set modeling of the rest, (iii) Synthesis result in two technologies, (iv) Power simulation of a prelayout netlist for single clock, (v) Extrapolation of the power consumption of single clock cycles for the cycle accurate simulation, (vi) Power values estimated from the cycle accurate simulation, and (vii) Area estimations after synthesis.

\begin{table}[t]
\caption{Total area consumption of 4-chiplet \papertitle 2.5D for different configurations on 28nm and 7nm.}
\centering
\begin{tabular}{l|c|c|c|c}
    \hline
     \textbf{Components}  & \multicolumn{2}{c|}{\textbf{28nm (mm$^2$)}}  & \multicolumn{2}{c}{\textbf{7nm (mm$^2$)}}\\ \cline{2-5}
      &  \textbf{1024$\times$64}   &   \textbf{512$\times$128}  & \textbf{1024$\times$64}   &   \textbf{512$\times$128} \\ \hline\hline
     \papertitle                  & 74.9  &  115  & 24 & 43.9 \\  \hline
      \papertitle-PU   & 58.0  &  81.0  & 7.01  & 9.9 \\
     $~~ \lfloor$       NTT/INTT  & 38.2  &  56.8  & 5.61  & 7.9 \\
     $~~ \lfloor$ 2$\times$MAS   & 3.1  &  6.6  & 0.42  & 0.76 \\
     $~~ \lfloor$       PRNG      & 0.15  &  0.28  & 0.02  & 0.04 \\
     $~~ \lfloor$ 2$\times$AUT    & 0.14  &  0.32  & 0.02  & 0.04 \\
     $~~ \lfloor$ Memory  & 16.1  &  16.1 &  1.2  & 1.2 \\ 
     HBM PHY/NoC          & 16.9  &  33.8  & 16.9  & 33.8\\\hline
     4$\times$\papertitle        & 299.6 &  392.4 & 96  &  175.6 \\
     C2C   & 12.32   &  14.64   &  0.8  & 1.6 \\\hline\hline
    \textbf{Total Area}          & \textbf{311.9} & \textbf{461.4} & \textbf{96.7} & \textbf{177} \\ \hline
\end{tabular}
\label{tab:area}
\end{table}

\autoref{tab:area} presents the area results for the \papertitle 2.5D architecture, featuring a 4-chiplet configuration as illustrated in \autoref{fig:reed_chiplet}. 
This design conforms to the fabricated chiplets systems~\cite{intel_xeon,amd_mi300}. The inner \papertitle-PU, NoC, and HBM (shown in \autoref{fig:reed_chiplet}) constitute one chiplet (similar to~\cite{occamy,manticore}). In \autoref{tab:perf}, we present the performance of FHE routines for both configurations (512$\times$128 and 1024$\times$64) with the achieved target clock frequency of 1.5 GHz. \autoref{sec:th_ks} explains how the throughput of KeySwitch is obtained. For $dnum=L+1$, we report high (${\approx}{99.9}\%$) utilization which reduces to ${\approx}{95}\%$ for $dnum=3$, as detailed in \autoref{sec:appendix_throughput}.

\subsection{Power and Performance Modelling}
Using a cycle-accurate model, we obtain the performance and power consumption estimates for \papertitle. \papertitle's communication and computation are decoupled by design and do not need application-specific schedules to reduce data load/store stalls. \papertitle also has a modular architecture design; all chiplets are identical, and none of the elementary building blocks is distributed across chiplets. The \papertitle-PU was fully described using Verilog Hardware Definition Language. Its complete functionality is tested via simulations in Vivado 2022.2. For the results, one REED-PU was fully synthesized to obtain overall area results using Cadence Genus and TSMC 28nm libraries. The entire system was modelled using identical instructions for software library and hardware implementation. However, in hardware, certain instructions can run simultaneously. Since they have constant execution time, we factored this behaviour into the software model, ensuring the clock cycle count aligned with expectations for hardware execution.

In~\autoref{sec:ISA}, we elaborated on how our instruction-set architecture handles the micro (NTT, AUT, MAS) and macro (e.g. rotation, KeySwitch) instructions. Thus, a user does not need to handle micro-instructions due to the provided macro-instruction level abstraction. Predefined microcode for static macro-instructions ensures optimized data flow and memory management. Data exchange across chiplets occurs solely during KeySwitch and is incorporated into the microcode and task distribution. The simulator takes into account the bandwidth of C2C and HBM-chiplet communication along with data communication and distribution strategies (\autoref{sec:data_dist},~\autoref{sec:non_blocking_comm}). The run-time of macro-instructions is obtained using the known static schedule of micro-instructions. Finally, the macro-instructions are scheduled using OpenFHE~\cite{openfhe}, and runtime is obtained for higher-level operations (bootstrapping, DNN).

We used the Cadence Genus tool to measure total power consumption for the entire \papertitle-PU. For each operation, the duration for which any unit in the chiplet remains active is predetermined and static.  The power consumption for each elementary building block is also obtained using the Cadence toolchain. We estimated the average power consumption based on this information combined with the runtime of each operation, following a methodology commonly employed by prior works such as ~\cite{ARK,BTS_isca,SHARP,CraterLake} for estimating FHE hardware acceleration on ASIC technology. Similarly, in a multi-chiplet scenario, the operation runtime remains static, and we conservatively estimate communication overhead to be $2\times$ slower than linear operations on the same volume of data. This approach provides a safe estimate of the total runtime. The cycle-accurate model serves as a simulator and effectively imitates the behaviour of the Instruction Set Controller within the chiplets, giving us a reasonable approximation of system performance. We validate this cycle-accurate modelling with Vivado simulation.

\subsection{What to expect from higher-throughput configurations?}
\pgfplotstableread[row sep=\\,col sep=&]{
    Work & Yield    \\
    16384$\times$4     & 75.6 \\
    8192$\times$8      & 80.6 \\
    4096$\times$16     & 82.6 \\
    2048$\times$32     & 87.6    \\
    1024$\times$64     & 96.7 \\
    512$\times$128     & 177 \\
    256$\times$256     & 332.4 \\
    128$\times$512     & 632.8 \\
    64$\times$1024     & 1216.6 \\
    }\areafirst

\pgfplotstableread[row sep=\\,col sep=&]{
    Work & Yield    \\
    16384$\times$4     & 18.9 \\
    8192$\times$8      & 19.9 \\
    4096$\times$16     & 20.9 \\
    2048$\times$32     & 21.9    \\
    1024$\times$64     & 24 \\
    512$\times$128     & 43.9 \\
    256$\times$256     & 82.6 \\
    128$\times$512     & 158.2 \\
    64$\times$1024     & 304.4 \\
    }\areafirstoncechiplet
\pgfplotstableread[row sep=\\,col sep=&]{
    Work & Yield    \\
    16384$\times$4     & 140.6 \\
    8192$\times$8      & 164.6 \\
    4096$\times$16     & 198.6 \\
    2048$\times$32     & 246.6 \\
    1024$\times$64     & 313.3 \\
    512$\times$128     & 474.4 \\
    256$\times$256     & 737.4 \\
    128$\times$512     & 1189.2 \\
    64$\times$1024     & 1983.8 \\
    }\areasecond
\pgfplotstableread[row sep=\\,col sep=&]{
    Work & Yield    \\
    16384$\times$4     & 31.9 \\
    8192$\times$8      & 37.6 \\
    4096$\times$16     & 46.9 \\
    2048$\times$32     & 58.9 \\
    1024$\times$64     & 74.9 \\
    512$\times$128     & 115 \\
    256$\times$256     & 181.6 \\
    128$\times$512     & 296.6 \\
    64$\times$1024     & 492.4 \\
    }\areasecondonechiplet
\begin{figure}[t]
        \centering
       \begin{tikzpicture}
    \begin{axis}[
            scale=0.9,
            enlarge x limits=0.02,
            height=230pt,
            width=\columnwidth,
            symbolic x coords={16384$\times$4,8192$\times$8,4096$\times$16,2048$\times$32,1024$\times$64,512$\times$128,256$\times$256,128$\times$512,64$\times$1024},
            xtick pos = bottom,
            xtick=data,
            ymax=320,
            xmin=16384$\times$4,
            xmax=64$\times$1024,
            x tick label style={rotate=45, anchor=east},
            grid style={line width=1pt, dotted,gray},
            extra y ticks={50,150,250,350}, 
            extra x ticks={512$\times$128,1024$\times$64},
            extra x tick style={grid=major,  grid style={dashed,blue}},
            extra y tick style={grid=major, grid style={dotted,gray}},
            every node near coord/.append style={rotate=-20, anchor=west ,opacity=0},
            xmajorgrids=true,
            ymajorgrids=true,
            xlabel={{Configuration ($N_1 \times N_2$)}},
            ylabel={{Area (mm$^2$)}},
            ylabel near ticks,
            every axis legend/.append style={ legend columns = 2}
        ]
         \addplot[thick, mark=square*,mark options={fill=blue!75!white},mark size=3.5pt,nodes near coords,scatter] table[x=Work,y=Yield ]{\areasecond};
         \addplot[thick, mark=diamond*,mark options={fill=cyan},mark size=5.2pt,nodes near coords,scatter] table[x=Work,y=Yield ]{\areasecondonechiplet}; 
        \addplot[thick, mark=triangle*,mark options={fill=orange!80!red},mark size=5pt,nodes near coords,scatter] table[x=Work,y=Yield ]{\areafirst};
        \addplot[thick, mark=*,mark options={fill=magenta},mark size=3.5pt,nodes near coords,scatter] table[x=Work,y=Yield ]{\areafirstoncechiplet};

         \addplot[thick,color=cyan,fill=cyan, 
                    fill opacity=0.1] coordinates  {
            (16384$\times$4, 50) 
            (16384$\times$4, 150) 
            (64$\times$1024,  150 )
            (64$\times$1024,  50 ) 
            (16384$\times$4, 50)
            };
            \addplot[thick,color=red,fill=red, 
                    fill opacity=0.1] coordinates  {
            (16384$\times$4, 40) 
            (16384$\times$4, 80) 
            (64$\times$1024,  80 )
            (64$\times$1024,  40 )  
            (16384$\times$4, 40) 
            };
            {\legend{monolithic with 4 \papertitle 28nm, 1 \papertitle chiplet 28nm, monolithic with 4 \papertitle 7nm, 1 \papertitle chiplet 7nm}}
            
    \end{axis}
\end{tikzpicture}
\caption{Area increase with rising throughput configurations \cite{chiplet_size_GPG}.}
        \label{fig:puarea}
    \end{figure}
Until now, we have examined two configurations (1024$\times$64 and 512$\times$128) that only partially demonstrate the advantages of our proposed scalable design methodology. As we double the throughput (by doubling the value of $N_2$), the area of PU only increases by approximately 1.5$\times$. This trade-off arises because the chip area comprises two components— $(i)$ the computation logic area, which scales linearly with throughput, and $(ii)$ on-chip storage that remains fixed to a number of polynomials. When we opt for a higher configuration, the polynomial size remains the same while the number of coefficients to be processed in parallel increases. 
We also discuss storing versus generating twiddle factors on the fly. As shown in the \autoref{tab:twidlle_comp}, twiddle factor generation (TFG) increases the number of multipliers (Mul.) only up to 18\% for different configurations ($N_1$ and $N_2$). Indeed, reduction in memory (Mem.) decreases (i.e., number of coefficients to store) with larger $N_2$. Yet, even for large values of $N_2$ (i.e., $N_1=64$ and $N_2=1024$), we observe a reduction in the memory by 57\% compared to storing the twiddle factors (\st{TFG}).

\begin{table}[t]
    \centering
    \resizebox{\columnwidth}{!}{
    \begin{tabular}{c|c|c|c|c|c|c}
       \hline $N_1\times N_2 \rightarrow$  & $2048\times32$ & $\textbf{1024}\times\textbf{64}$ & $512\times128$ & $256\times256$ & $128\times512$ & $64\times1024$    \\ \hline\hline
       \textbf{Total Mul.}  & 432 & \textbf{832} & 1,600 & 3,072 & 5,888 & 11,264\\ \hline \hline 
       \textbf{TFG Mul.}  & 68 & \textbf{131} & 258 & 513 & 1,024 & 2,047 \\ 
       \textbf{TFG Mem.}  & 222,912 & \textbf{310,624} & 486,400 & 707,232 & 1,149,248 & 1,771,488 \\ \hline\hline
       \textbf{\st{TFG} Mul.}  & 0& \textbf{0} & 0 & 0 & 0 & 0 \\ 
       \textbf{\st{TFG} Mem.}  &  4,260,320 & \textbf{4,228,064} & 4,212,704 & 4,206,560 & 4,206,560, & 4,212,704 \\ 
       \hline\hline\textbf{Mul. Inc. }  & 16\% & \textbf{16\%} & 16\% & 17\% & 17\% & 18\% \\ 
       \textbf{Mem. Red.}  & 95\% & \textbf{93\%} & 88\% & 83\% & 73\% & 58\% \\ \hline
    \end{tabular}
    }
    \caption{The table compares memory (number of coefficients to store) and multiplier units for different $N_1$ and $N_2$ values with and without twiddle factor generation (TFG).}
    \label{tab:twidlle_comp}
\end{table}

However, an important question remains: \textit{what configuration strikes the best balance between throughput and manufacturing cost?}  To address this, we turn to~\cite{chiplet_size_GPG}. The authors report that the best manufacturing size for high yield ranges from 40 to 80 mm$^2$ for 7nm technology, while for 40nm, it ranges from 50 to 150 mm$^2$. In \autoref{fig:puarea}, we present two sets of area consumption results for 28nm and 7nm technologies. The first set corresponds to four \papertitle cores produced as a single monolithic chip, while the second set represents one \papertitle chiplet. The best area ranges are highlighted in black and pink. As we can see, for both 7nm and 28nm, the configuration 512$\times$128 falls within the best development area range and offers high throughput. The configuration 1024$\times$64 is within the optimum range for 28nm and is close to it for 7nm. Monolithic designs, within the best range, offer 4$\times$ to 8$\times$ less throughput.

\subsection{Comparison with Related Works}

The realization of privacy-preserving computation through FHE holds great potential for the entire community, resulting in various acceleration works. Among these, the ASIC designs~\cite{feldmann_2021f1,ARK,BTS_isca,CraterLake,SHARP} have achieved the most promising acceleration results. However, a direct comparison with these works would be unfair as the benchmarks are provided for different parameters ($dnum,L,w,L_{boot}$). Hence, to ensure fairness, we also provide bootstrapping results for $dnum=3$ (utilized by~\cite{SHARP}) in~\autoref{tab:comp}. Next, since we cannot change the word size ($w=54$) chosen for high precision, we select  $L=23, K=8, \text{ and } L_{boot}=17$ accordingly. 

We use the amortized bootstrapping time T$_{\text{A.S.}}$~\cite{BTS_isca,ARK} metric that calculates the bootstrapping time divided by $L_{\textit{eff}}$ and packing $n$. This metric overlooks factors such as area, power, and precision.  Higher precision necessitates a larger word size, $w$ (or expensive composite scaling).  Thus, we use the EDAP (Energy-Delay-Area product) metric~\cite{edap} and modify it (EDAP$_w$) to incorporate a linear
increase due to word size (discussed in \autoref{sec:prec_loss}).

\autoref{tab:comp} compares our design's area consumption, performance, and power consumption for the packed bootstrapping operation with existing monolithic works, F1~\cite{feldmann_2021f1}, BTS~\cite{BTS_isca}, ARK~\cite{ARK}, CraterLake (CLake)~\cite{CraterLake}, and SHARP (SH) \cite{SHARP}. \papertitle achieves $1.9\times$ better performance than the state-of-the-art (SH$_{36}$) while consuming $1.8\times$ less area. Our area consumption is less as the prior works utilize at least half of the chip area for on-chip memory. In our case, on-chip memory is not significant, as we utilize HBMs for major storage. We obtain better performance results due to the high throughput and $4.8\times$ higher off-chip communication bandwidth offered by four HBM3 blocks -- where each chiplet is exclusively connected to one HBM3.

\begin{table}[t]

\caption{Comparison of \papertitle 2.5D  with state-of-the-art}\label{tab:comp}
\centering
\begin{tabular}{l|c|c|c|c|c|c}
    \hline
    \multirow{2}{*}{\textbf{Work}} & \textbf{Area}  & \textbf{ T$_{\text{A.S.}}$}  & \textbf{P$_{\text{Avg}}$} & \textbf{EDAP$_w$} & \textbf{Parameters} & \textbf{\# B.S.}/Dollar \\
       &  (mm$^2$)  & (ns) & (W) &  (/M)   & ($N/L/dnum$) &(ops)$^*$~\cite{chiplet_size_GPG}\\\hline\hline
    F1$_{32}$~                & 71.02$^{\dag}$  & 470  & 28.5$^{\dag}$   & 754.5 & $2^{14}/23/24$ & 0.48\\
    BTS$_{64}$                & 373.6      & 45.4 & 163.2& 106.0 & $2^{17}/39/2$ & 0.74\\
    ARK$_{64}$               & 418.3      & 14.3 & 135  & 9.74 & $2^{16}/23/4$ & 1.95 \\
    CLake$_{28}$         & 222.7$^{\dag}$  & 17.6 & 124$^{\dag}$  & 16.5 & $2^{16}/60/{1 \hspace{-.25em}-\hspace{-.25em}3}$ & 3.64\\
    SH$_{36}$       & 178.8      & 12.8 & 94.7 & 4.1 & $2^{16}/35/3$  & 6.23\\
    SH$_{64}$        & 325.4     & 11.7 & 187  & 7.0 & $2^{16}/22/3$ & 3.4 \\ \hline\hline
    
    \multirow{2}{*}{\textbf{\papertitle}$_{54}^{\ddag(1)}$} &\textbf{98.4$^{**}$}      & \textbf{6.6} & \textbf{48.8} & \textbf{0.21} & $2^{16}/23/3$ & \textbf{25.3}\\ \cline{2-7}
     &   \textbf{96.7}   & \textbf{28.8} &  \textbf{49.4} & \textbf{3.96} & $2^{16}/30/31$ & \textbf{6.32}\\ \hline
     \textbf{\papertitle}$_{54}^{\ddag(2)}$ &  \textbf{177} & \textbf{14.4} & \textbf{83.5} & \textbf{3.10} & $2^{16}/30/31$ & \textbf{6.82}\\
     \hline

\end{tabular}
\begin{flushleft}
$^{\dag}$ Area/power are normalized~\cite{area_scale,SHARP} (14nm/12nm to 7nm).\\
$^{\ddag}$ Result for configuration (1) $1024\times64$ with one HBM per chiplet,  and (2) $512\times128$ with two HBM per chiplet.\\
$^{*}$ The cost per mm$^2$ is obtained from~\cite{pricing} ($\$57,500/$mm$^2$ for 7nm) and yield estimates are taken from~\cite{chiplet_size_GPG,chiplet_size_MaWWCH22} including manufacturing, pre- and post-testing, and integration costs.\\
$^{**}$ For $dnum<L+1$, the area increases due to BaseConversion MAS units (\autoref{sec:appendix_throughput}).
\end{flushleft}
\end{table}

\pgfplotstableread[row sep=\\,col sep=&]{
    Work & ManufacturingCost/yield    \\
    F1    & 1.2\\
    BTS    & 2.2\\
    CLake    & 1.5\\
    ARK    & 2.5\\
    SH$_{36}$    & 1.5\\
    SH$_{64}$    & 2.0\\
    RD$_{_{7nm}}$    & 1\\
    RD$_{_{28nm}}$ & 0.3 \\
    }\costyield 

\pgfplotstableread[row sep=\\,col sep=&]{
    Work & Yield    \\
    F1    & 0.95 \\
    BTS     & 0.74 \\
    CLake     & 0.85 \\
    ARK      & 0.69 \\
    SH$_{36}$    &  0.87 \\
    SH$_{64}$    & 0.76 \\
    RD$_{_{7nm}}$  & 1 \\
    RD$_{_{28nm}}$  & 0.98 \\
    }\yield
\pgfplotstableread[row sep=\\,col sep=&]{
    Work & Cost    \\
    F1     & 1.1 \\
    BTS     & 1.6 \\
    CLake     & 1.3 \\
    ARK      & 1.7 \\
    SH$_{36}$    & 1.3 \\
    SH$_{64}$    & 1.5 \\
    RD$_{_{7nm}}$    & 1 \\
    RD$_{_{28nm}}$  & 0.3 \\
    }\cost
    
\begin{scriptsize}
\begin{figure}[]
       \centering
       \begin{tikzpicture}
    \begin{axis}[
            ybar=0.0cm,
            width=1.05\columnwidth,
            height=210pt,
            symbolic x coords={F1,BTS,CLake,ARK,SH$_{36}$,SH$_{64}$,RD$_{_{7nm}}$,RD$_{_{28nm}}$},
            ymax=3.4,
            xtick pos = bottom,
            bar width=12pt,
            nodes near coords,
            every node near coord/.append style={rotate=90, anchor=west},
            axis background/.style={fill=white!5},
            xtick=data,
            extra x ticks={RD$_{_{7nm}}$,RD$_{_{28nm}}$}, 
            extra y ticks={1}, 
            extra y tick style={grid=major, grid style={dashed,red}},
            grid style={line width=1pt, dotted,gray},
            ymajorgrids=true,
            every axis legend/.append style={ legend columns = 3},
            ylabel={{Relative value} },
            ylabel near ticks
        ]
        \addplot[fill=brown!40!yellow,draw=blue!20!black] table[x=Work,y=Yield]{\yield};
        \addplot[fill=cyan,draw=cyan!10!black] table[x=Work,y=Cost]{\cost};
        \addplot[black,fill=blue!20!white,  postaction={pattern=north east lines}] table[x=Work,y=ManufacturingCost/yield ]{\costyield};        
        {\legend{{Relative Yield}, {Relative Cost}, {Relative Cost/yield}}}
    \end{axis}
\end{tikzpicture}
\caption{Relative a) yield of existing monolithic designs versus the proposed 7nm chiplet-based architecture \cite{chiplet_size_MaWWCH22}, b) development cost (including Interposer cost) \cite{chiplet_size_GPG,pricing,medha}, and c) cost of SiP development (cost/yield). RD refers to our work \papertitle 2.5D. }
    \label{fig:cost_yield}
    \end{figure}
\end{scriptsize}

Prior works utilize up to $1$TB/s peak off-chip bandwidth, relying on large on-chip memories for storing keys, which necessitates much higher on-chip communication bandwidths— $20$TB/s~\cite{ARK}, $36$TB/s~\cite{SHARP}, and $84$TB/s~\cite{CraterLake}. In contrast, although our proposed \papertitle system uses $4.8\times$ higher off-chip bandwidth (with each HBM offering $1.2$TB/s at its peak), it efficiently offloads memory requirements to HBM, resulting in significantly reduced on-chip memory usage and consequently $4.2-7.5\times$ lower on-chip memory bandwidth requirements compared to~\cite{ARK,SHARP,CraterLake}. The off-chip bandwidth requirement of one \papertitle chiplet is comparable to that of a monolithic design. However, it incurs a $2.7\times$ performance loss compared to the state-of-the-art~\cite{SHARP} while consuming $3.36\times$ less area. This trade-off is a consequence of designing chiplet-based architecture over a monolithic one. The \papertitle 4-chiplet system offers better performance and $35\times$ higher energy efficiency for lower chiplet area, as shown in~\autoref{tab:comp}.

We also assess the yield and manufacturing cost in \autoref{fig:cost_yield} by utilizing the results reported in \cite{chiplet_size_MaWWCH22,chiplet_size_GPG,pricing} including manufacturing, pre- and post-testing, and integration costs. For a fair comparison, we use the original area and not the word-size scaled area for prior works. As illustrated, we achieve the highest yield and lowest manufacturing cost on 7nm, resulting in the least overall cost (manufacturing cost/yield), 50\% less than state-of-the-art monolithic design SHARP$_{64}$. The cost is estimated using the yield metric provided in \cite{chiplet_size_MaWWCH22,FengM22}. Based on this analysis, we observe that the yield for SHARP$_{64}$ is approximately $73\%$, whereas the yield for the 4-chiplet \papertitle ($512 \times 128$) configuration reaches around $96\%$. This represents a $1.31\times$ improvement in relative yield. Factoring in the manufacturing costs\cite{pricing} and the chiplet packaging costs~\cite{chiplet_size_GPG}, the design achieves an additional $1.5\times$ cost reduction. Together, these factors contribute to a $50\%$ reduction in the cost-to-yield ratio, offering an estimate of the overall chip manufacturing cost. On 28nm technology, we achieve $85\%$ cheaper design compared to SHARP$_{64}$. 
Further, note that the cost per bootstrapping for $dnum=L+1$ looks similar to prior monolithic works, but at this parameter, we also have the highest ($\approx2\times$ more) computation depth remaining after bootstrapping for $\omega\geq 54$. Thus, applications utilizing this parameter choice require $2\times$ less frequent bootstrapping operations.

\subsection{Higher Chiplet-Integration Study}\label{sec:higherchiplet}

In \autoref{sec:mc_des}, we examined multiple chiplet configurations and selected four ($\mathtt{r}=4$) based on long-term utilization, lower power dissipation, and low integration costs. Our design methodology and data/task distribution approach remain adaptable to any desired chiplet configuration and the number of chiplets.  In fully connected chiplet nodes, interconnect length between chiplets can significantly impact energy consumption and latency. Therefore, our proposed non-blocking ring-based communication technique for KeySwitching shows a better advantage for higher chiplet integration density. 

\begin{figure}[t]
    \centering
    \includegraphics[width=0.55\columnwidth]{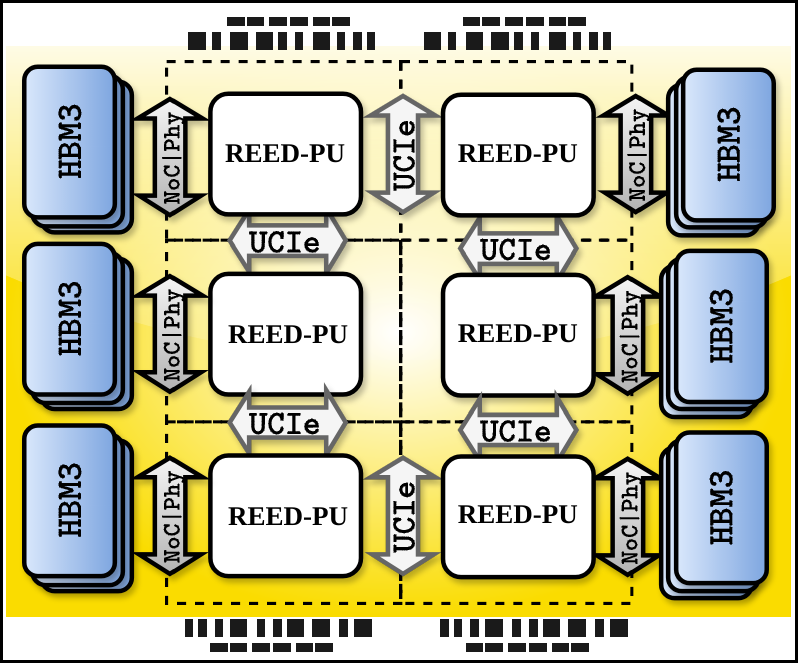}
    \caption{The complete architecture diagram of 6-chiplet \papertitle 2.5D for 1024$\times$64 configuration.}
    \label{fig:six_chip}
\end{figure}

\begin{table}[h]
    \centering
    \begin{tabular}{c|c|c|c}
    \hline
         \textbf{Number of Chiplets} $\rightarrow$ & \textbf{4} & \textbf{8} & \textbf{12}  \\ \hline\hline
         \textbf{Area} ($mm^2$) & 311.9 & 623.8 & 935.7 (> reticle limit)\\ \hline
         \textbf{T$_{\text{A.S.}}$} ($ns$) & 14.41 & 7.37 & 4.98\\\hline
    \end{tabular}
    \caption{ Results for $1024\times64$ configuration on 28nm technology with parameters ($N/L/dnum=2^{16}/30/31$).}
    \label{tab:integration}
\end{table}

\autoref{fig:six_chip} illustrates a chiplet-based architecture for six chiplets, which can be expanded to accommodate more chiplets. Additionally, energy-efficient lower-bandwidth memories, such as DDR, can be integrated with the appropriate ($N_1, N_2$) configuration based on memory throughput. While increasing the number of chiplets enhances performance, as shown in Table~\ref{tab:integration}, it comes at the cost of area and underutilization when the current multiplicative depth ($l$) falls below the number of chiplets ($l<r$). However, more chiplets can lead to better performance in the long term and allow tiling beyond the reticle limit, albeit with the additional area, power, and integration overhead. Also, note that a higher number of chiplet interconnects implies additional points of failure, making testability and reliability more involved. A more detailed multi-chiplet study is also presented in concurrent work \cite{KimKCA24}, where the authors show that increasing the cores from 4 to 16 results in speedup of only $1.99-2.11\times$.

\section{Application benchmarks}\label{sec:application_benchmarks}

We benchmark three machine learning applications: linear regression, logistic regression, and a Deep Neural Network (DNN). Each application is evaluated for \textit{encrypted} training and inference. In this setting, the server provides computational support without knowledge of the data or model parameters, ensuring complete blind computation. Most applications benchmarked in the previous works~\cite{feldmann_2021f1} are partially blind; the server does not see the data but knows the model parameters to evaluate it. To the best of our knowledge, none of the previous works benchmark an encrypted neural network training. The speedup results are presented in \autoref{tab:appl_speedup} using the most area conservative design (1024$\times$64). Higher configuration (512$\times$128) will improve the performance by 2$\times$.

\begin{table}[t]
    \centering
    \caption{Application benchmark and the speedup achieved by \papertitle 2.5D. The CPU speed is reported on a 24-core, 2$\times$Intel Xeon CPU X5690 $@$ 3.47GHz with 192GB DDR3 RAM. }
    \begin{tabular}{c|c|c|c|c|c}
    \hline
         \multirow{2}{*}{\textbf{Appl.}} &   \multirow{2}{*}{\textbf{Accuracy}} & \multirow{2}{*}{\textbf{Op}} &\multicolumn{2}{c|}{\textbf{Time}} & \multirow{2}{*}{\textbf{Speedup}} \\ \cline{4-5}
          &   &  &\textbf{ CPU} & \textbf{HW } & \\ 
         \hline\hline
         \multirow{2}{*}{Lin.Reg. }& \multirow{2}{*}{ 78.12\%}& Inf. & 0.86 s &  0.31 ms & 2,873$\times$\\ \cline{3-6}
         & & Trn.   & 13.82 s & 4.6 ms& 2,991$\times$\\\hline
         \multirow{2}{*}{Log.Reg.} & \multirow{2}{*}{61.8\%} & Inf. &  1.27 s &  0.46 ms & 2,785$\times$\\ \cline{3-6}
         & & Trn.  &  11.18 s & 3.8 ms  & 2,865$\times$\\\hline
         \multirow{2}{*}{DNN }& \multirow{2}{*}{95.2\%} & Inf. & 128.7 s& 48.6 ms& 2,646$\times$ \\\cline{3-6}
         & & Trn.  & 29 days & 920 s & 2,725$\times$\\\hline
    \end{tabular} 
    \label{tab:appl_speedup}
\end{table}

\pgfplotstableread[row sep=\\,col sep=&]{
    Work         & EDAP*    \\
    SH$_{36}$    & 1.6 \\
    CLake$_{28}$ & 8.7 \\
    SH$_{64}$    & 3.3 \\
    REED$_{54}$   & 1 \\
    }\speedupEDAPone
\pgfplotstableread[row sep=\\,col sep=&]{
    Work         & EDAP/w    \\
    CLake$_{28}$        & 10.9 \\
    SH$_{36}$    & 1.2 \\
    SH$_{64}$    & 3.7 \\
    REED$_{54}$   & 1 \\
    }\speedupEDAPwone 
\pgfplotstableread[row sep=\\,col sep=&]{
    Work         & EDAP*    \\
    SH$_{36}$    & 1.6 \\
    REED$_{54}$   & 1 \\
    }\speedupEDAPtwo
\pgfplotstableread[row sep=\\,col sep=&]{
    Work         & EDAP/w    \\
    SH$_{36}$    & 1.3 \\
    REED$_{54}$   & 1 \\
    }\speedupEDAPwtwo
\begin{scriptsize}
\begin{figure}[t]
       \centering
       \begin{tikzpicture}
       \centering
    \begin{axis}[
    colormap/bluered,
            width=0.7\columnwidth,
            height=200pt,
            ybar=0.01cm,
            enlarge x limits=0.2,
            symbolic x coords={CLake$_{28}$,SH$_{36}$,SH$_{64}$,REED$_{54}$},
            ymax=11.2,
            ymin=0.01,
            xtick pos = bottom,
            bar width=12pt,
            nodes near coords,
            every node near coord/.append style={thick,rotate=90, anchor=east},
            ylabel={Relative value},
            axis background/.style={fill=white!5},
            xtick=data,
            extra y ticks={1},
            extra x ticks={REED$_{54}$},
            extra y tick style={grid=major, grid style={dashed,red}},
            grid style={line width=1pt, dotted,gray},
            ymajorgrids=true,
            every axis legend/.append style={ legend columns = 1}
        ]
        \addplot[fill=white!35!gray,draw=black] table[x=Work,y=EDAP/w]{\speedupEDAPwone};
        \addplot[fill=white!15!olive!15!orange,draw=black] table[x=Work,y=EDAP*]{\speedupEDAPone};
        \addplot[fill=white!70!gray,draw=black] table[x=Work,y=EDAP/w]{\speedupEDAPwtwo};
        \addplot[fill=white!15!cyan,draw=black] table[x=Work,y=EDAP*]{\speedupEDAPtwo};
        {\legend{Relative EDAP$_w$ [HELR256],Relative EDAP$_{w^2}$[HELR256], Relative EDAP$_w$[HELR1024], Relative EDAP$_{w^2}$[HELR1024]}}
    \end{axis}
    
\end{tikzpicture}
\caption{Relative metrics comparison for the HELR \cite{helr} application with batch sizes 256 and 1024. Under these metrics, the lower the value, the better.}
        \label{fig:compare_speedup}
    \end{figure}
\end{scriptsize}

\begin{itemize}
    \item \textbf{Linear Regression}: We employ the Kaggle Insurance dataset~\cite{lin_reg} to benchmark linear regression. The model uses a batch size of 1204 and 1338 input feature vectors (each containing six features) for training and inference and achieves an accuracy of 78.1\% (same as plain model~\cite{lin_reg}).

\item  \textbf{Logistic Regression}: It is a supervised machine learning model that utilizes the log function, evaluated using function approximations in a homomorphic context. Its accuracy depends on the degree of approximation function expansion and precision.  Existing works, such as~\cite{CraterLake,SHARP}, utilize the HELR~\cite{helr} to benchmark encrypted training on MNIST~\cite{mnist} data, with batch sizes (256, 1024). In \autoref{fig:compare_speedup}, we illustrate the performance advantage of \papertitle 2.5D. To predict cancer probability, we further evaluate logistic regression on the iDASH2017 cancer dataset (similar to~\cite{KSKLC18}). Here, we achieve a training accuracy of 62\% in a single iteration. This dataset comprises 18 features per input, with batch sizes of 1422 and 1579 used for training and inference.

\begin{figure}[t]
    \centering
    \includegraphics[width=0.45\columnwidth,trim={1cm 0.1cm 0 0},clip]{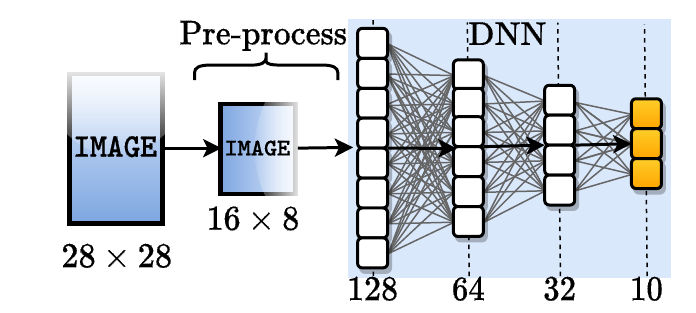}
    \caption{A DNN  for MNIST~\cite{mnist} with two hidden and one output layers.}
    \label{fig:dnn}
\end{figure}

\item \textbf{Deep Neural Network }: The DNN serves as a powerful tool for Deep Learning. Our study employs a DNN, shown in \autoref{fig:dnn}, for the MNIST dataset~\cite{mnist}, with two hidden and one output layer. We pack four pre-processed images per batch to prevent overflow during matrix multiplication. DNN training requires 12,500 batches. Thus, all the existing works~\cite{ARK,BTS_isca,CraterLake,SHARP} not providing computation-communication parallelism will suffer as their on-chip memory is insufficient. The DNN is trained for $\approx$7000 ($\approx$5.8 Bootstrappings per iteration) iterations and achieves 95.2\% accuracy in 29 days using OpenFHE~\cite{openfhe}. \papertitle 2.5D could finish this in only 15.4 minutes. This is where our computation-communication parallelism shines, as many ciphertexts are required for such an application. None of the works in literature offers this and is bound to suffer for memory-intensive applications. 

\end{itemize}

\begin{figure}[t]
    \centering
    \includegraphics[width=0.8\columnwidth,trim={0.1cm 0 0.1cm 0},clip]{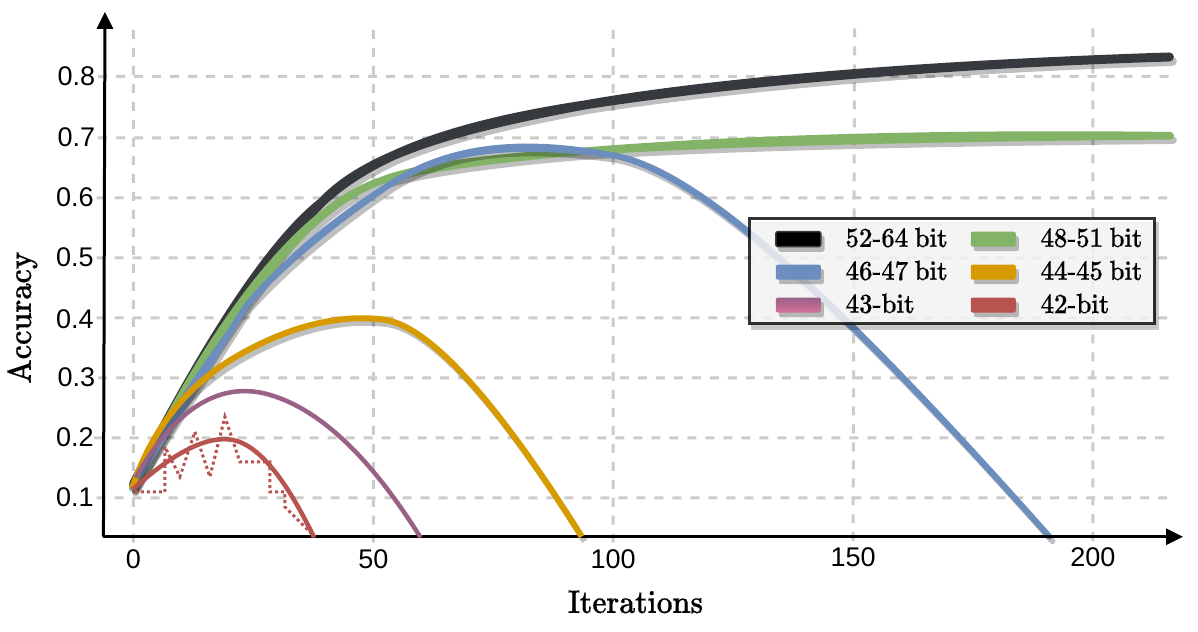}
    \caption{Accuracy plot of different word sizes for the DNN. The lines are smoothened and the red dotted zig-zag line resembles the original form.}
    \label{fig:prec_loss_dnn}
\end{figure}

\subsection{Precision-loss Experimental Study}\label{sec:prec_loss} 
Another facet of privacy-preserving computation is precision loss. Since the server cannot see the intermediate or final results, the best it can do is to ensure that the parameters it operates on support higher precision. To validate our parameter sets, we ran experiments for the DNN training. In \autoref{fig:prec_loss_dnn}, we can see how quickly the training accuracy drops as the word size is reduced. Thus, precision plays a vital role in providing privacy-preserving computation on the cloud. Our choice of 54-bit word size strikes the perfect balance between precision and performance. Works offering a smaller word-size~\cite{feldmann_2021f1,CraterLake,SHARP} require in-depth study to mitigate the accuracy loss due to low precision.


\section{Future Scope: Journey from 2.5D  to 3D}\label{sec:3dchiplet}

\begin{figure}[t]
    \centering
    \includegraphics[width=0.7\columnwidth]{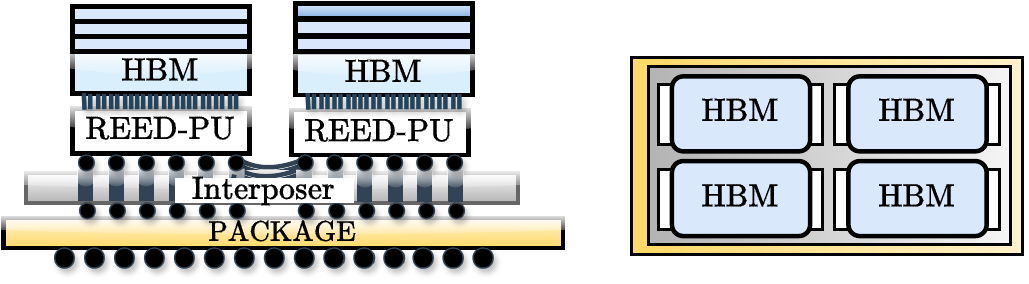}
    \caption{The side and top view of futuristic \papertitlemain has four \papertitle 3DIC chiplets. }
    \label{fig:3dreed}
\end{figure}

The extension of \papertitle 2.5D to a complete 3D IC holds immense potential for future computing. To achieve this transition, we have two options: connecting the PU with the HBM controller via TSV (as shown in \autoref{fig:3dreed}) or merging the PU unit with the lower HBM controller die. Since HBM is sold as an IP, the latter approach relies on the IP vendors to integrate the PU. By adopting either of these approaches, we can significantly reduce the reliance on the Network-on-Chip (NoC), leading to a compact chip design with lower power consumption. Each chiplet will be a full 3D IC package (PU and Memory) and will need a C2C link via interposer for connecting to other chiplets. A reduction in the area is expected due to fewer HBM stacks on the lateral area and the integration of the \papertitle-PU unit with the HBM controller. Additionally, decreased critical paths would further enhance the design's performance. Thus, the \papertitle's 3D IC integration promises a huge reduction in overall chip area and power consumption.

\section{Conclusion}\label{sec:disc_and_conc}

FHE has garnered considerable interest due to its ability to preserve data and computation privacy, leading to several efforts for accelerating FHE using large monolithic ASIC designs. However, many of these attempts primarily focus on acceleration at the expense of yield, manufacturing cost, and scalability. We proposed a scalable FHE accelerator design methodology for a multi-chiplet system that can be easily extended to larger configurations while adapting to constrained environments.

Chiplet-based designs face inherent challenges, such as increased latency costs due to slow C2C communication. We designed \papertitle to address these and show advantages over the monolithic designs in terms of performance, area, and energy consumption. \papertitle achieved this feat by utilizing a non-trivial yet uncomplicated bandwidth-oriented design methodology and a modular design approach.

An efficient workload division strategy optimized for multi-chiplet architectures is proposed to minimize memory usage and enhance efficiency through interleaved data and workload distribution strategy for all FHE routines. Next, to address slow chiplet-to-chiplet communication, a novel non-blocking ring-based inter-chiplet communication strategy tailored to FHE is introduced. Additionally, a scalable bandwidth-oriented design methodology is adopted, offering flexibility to adjust to the varying area and performance needs while supporting communication-computation parallelism within each chiplet. Furthermore, novel design techniques are presented for building blocks (NTT/AS/AUT), enhancing scalability. These techniques accelerate routines such as KeySwitch or Bootstrapping and reduce their latency by ${\approx}{67\%}$.  Finally, \papertitle benchmarks an encrypted DNN training, demonstrating utility for real-world FHE applications.

All the \papertitle-chiplets are small and identical, allowing us to prototype the building blocks using FPGA and validate the functionality, which had not been done by prior works. Summarily, the proposed design methodology for \papertitle showcased a robust and scalable approach to FHE acceleration that addressed several key challenges inherent in traditional monolithic designs. This paves the way for interesting future prospects such as formal verification. The advancements presented in this work hold the promise of advancing privacy-preserving computations and promoting the wider adoption of fully homomorphic encryption.

\section*{Acknowledgement}
This work was supported in part by Samsung Electronics Co. Ltd., Samsung Advanced
Institute of Technology and the State Government of Styria, Austria – Department
Zukunftsfonds Steiermark. We extend our gratitude to the anonymous reviewers for their constructive feedback. We thank Ian Khodachenko for conducting extensive application benchmarking, the results of which were integral to this paper, and Florian Hirner for FPGA prototyping.


\appendix
\section*{Appendix}

\section{KeySwitch Task Distribution for \text{$dnum<L+1$ ($k>1$)}}\label{appendix:keyswitch}
    \begin{algorithm}[h]
\renewcommand{\algorithmicrequire}{\textbf{In:}}
\renewcommand{\algorithmicensure}{\textbf{Out:}}
\caption{$\mathtt{CKKS.KeySwitch}$ ~\cite{HK20,KimPP22} (for arbitrary $dnum$)}
\label{algo:relin_gen}
\begin{flushleft}
\textbf{In:}  $\mathtt{d} = (\mathbf{\tilde{{d}}}_{0}, \mathbf{\tilde{{d}}}_{1}, \mathbf{\tilde{{d}}}_{2}) \in R_{Q_l}^{3}$, $\tilde{\mathtt{ksk}}_0 \in R_{PQ_{l}}^{dnum}, \tilde{\mathtt{ksk}}_1 \in R_{PQ_{l}}^{dnum}$ 

 \textbf{Out:} $\mathtt{d}' = (\mathbf{\tilde{{d}}}'_{0}, \mathbf{\tilde{{d}}}'_{1}) \in R_{Q_l}^{2}$ 
\end{flushleft}
\vspace{-10pt}
\begin{algorithmic}[1]
    \FOR{$j=0$ to $dnum-1$}	
	\STATE $\mathbf{\tilde{y}}[j] \leftarrow {\tilde{{d}}}_{2}[j\cdot K:(j+1)\cdot K-1] ~\cup~ \mathtt{BConvRout}_{Q_K\rightarrow PQ_{l/K}}({\tilde{{d}}}_{2}[j\cdot K:(j+1)\cdot K-1])$
	    \STATE $(\mathbf{\tilde{{c}}}''_{0}[j], \mathbf{\tilde{{c}}}''_{1}[j]) \leftarrow 0$
	    \FOR{$i=0$ to $l+K$}	
    	\STATE ${\tilde{{c}}}''_{0}[j][i] \leftarrow \big[{\tilde{{c}}}''_{0}[j][i] + \tilde{\mathtt{ksk}}_{0}[j][i] \cdot {\tilde{{y}}[j][i]}\big]_{q_j}$\STATE  ${\tilde{{c}}}''_{1}[j][i] \leftarrow \big[{\tilde{{c}}}''_{1}[j][i] + \tilde{\mathtt{ksk}}_{1}[j][i] \cdot {\tilde{{y}}[j][i]}\big]_{q_j}$
\ENDFOR	
	\ENDFOR	
     \FOR{$j=1$ to $dnum-1$}	
    	\STATE ${\mathbf{\tilde{{c}}}}''_{0}[0] \leftarrow \big[{\mathbf{\tilde{{c}}}}''_{0}[0] +{\mathbf{\tilde{{c}}}}''_{0}[j]]$
     \STATE ${\mathbf{\tilde{{c}}}}''_{1}[0] \leftarrow \big[{\mathbf{\tilde{{c}}}}''_{1}[0] +{\mathbf{\tilde{{c}}}}''_{0}[j]]$  
	\ENDFOR	
    \STATE $\mathbf{\tilde{{d}}}'_{0} \leftarrow \mathbf{\tilde{{d}}}_{0} + (\mathbf{\tilde{{c}}}''_{0}[0]_{Q_l}- \mathtt{BConvRout}_{P\rightarrow Q_l}(\mathbf{\tilde{{c}}}''_{0}[0])$)
    \STATE $\mathbf{\tilde{{d}}}'_{1} \leftarrow \mathbf{\tilde{{d}}}_{1} + ( \mathbf{\tilde{{c}}}''_{1}[0]_{Q_l} - \mathtt{BConvRout}_{P\rightarrow Q_l}(\mathbf{\tilde{{c}}}''_{1}[0]))$   
\end{algorithmic}
\end{algorithm}

    \begin{algorithm}[t]
\renewcommand{\algorithmicrequire}{\textbf{In:}}
\renewcommand{\algorithmicensure}{\textbf{Out:}}
\caption{$\mathtt{CKKS.BconvRout}_{P\rightarrow Q_l}$}
\label{algo:bconv_gen}
\begin{flushleft}
\textbf{In:}  $ \mathbf{\tilde{{y}}}_{2} \in R_{P}$

 \textbf{Out:} $\mathbf{\tilde{{y}}} \in R_{Q_l}$ 
\end{flushleft}
\vspace{-12pt}
\begin{algorithmic}[1]
\FOR{$j=0$ to $K-1$}	
	\STATE $y_2[j]_{p_j} \leftarrow \mathtt{INTT}({\tilde{{y}}}_{2}[j])$
    \ENDFOR
\FOR{$j=0$ to $l$}	
    \STATE $y[j]_{q_j} \leftarrow \sum^{K-1}_{i=0} (y_2[i]\cdot \hat{p}^{-1}_i \bmod{p_i})\cdot \hat{p_i} \bmod{q_j}$ \hfill \textcolor{gray}{$\triangleright$ BaseConversion}
	\STATE $\tilde{y}[j]_{q_j} \leftarrow \mathtt{NTT}({{{y}}}[j+K])_{q_j}$
    \ENDFOR  
\end{algorithmic}
\end{algorithm} 

The generic version of the KeySwitch routine~\cite{HK20,KimPP22} is presented in \autoref{algo:relin_gen}. The $L+1$ limbs of the ciphertext are split into $dnum$ digits of $K$ limbs. Every digit consisting of $K$ limbs is used to obtain $l+K+1$ limbs after the ModUp operation (via BConvRout \autoref{algo:bconv_gen}) for ciphertext at depth $l$. This results in $dnum\cdot (l+K+1)$ limbs, which are then multiplied with the keys and accumulated to return  $(l+K+1)$ limbs per ciphertext component. Finally, a ModDown is done to reduce the limbs to  $(l+1)$ limbs and the operation flow is very similar to that of ModUp. As discussed in \autoref{sec:lc_dist}, maintaining the limb-based decomposition proves advantageous compared to switching between limb-based and coefficient-based methods. We now explore how this limb-based decomposition applies when $dnum<L+1$. It is important to note that for $dnum<L+1$, NTT computation cannot start until the INTT results are multiplied with base hats ($\hat{p_i},\hat{p_i}^{-1}$) for BaseConversion.

\begin{figure}[t]
    \centering
    \includegraphics[width=\linewidth,trim={0.8cm 0.3cm 2.2cm 0.2cm},clip]{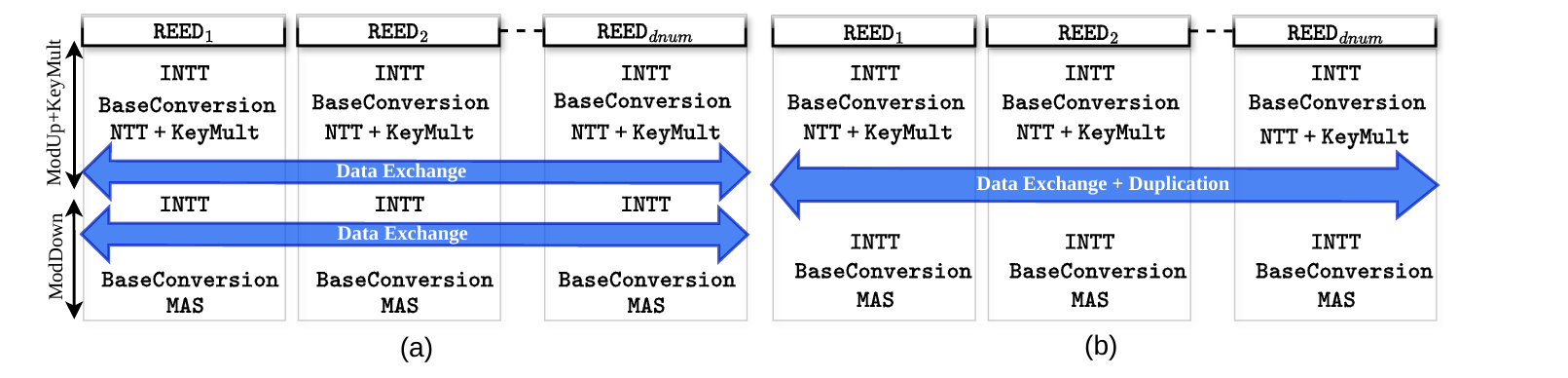}
    \caption{The Data and Task distribution when limbs are distributed digit-wise for KeySwitch computation. The INTT, NTT, and BaseConversion mentioned here refer to computations required per digit. In the first case, Data Exchange is done in parallel with BaseConversion, and in the second case, it is done in parallel with KeyMult. This does not produce a full decoupling effect when C2C communication is slower.}
    \label{fig:exch_1}
\end{figure}

The first technique involves digit-wise distributing data and tasks across limbs, where the number of chiplets equals $dnum$. This allows each chiplet to independently execute the outer loop for ModUp in \autoref{algo:relin_gen} without requiring inter-chiplet data exchange. Data is only shared after ModUp (before Key Multiplication) and during ModDown, as illustrated in \autoref{fig:exch_1} first figure. The authors in~\cite{ARK}, note that this has an overhead of $2 \cdot dnum \cdot (K + l + 1)$ polynomials per chiplet, which can be reduced to $\frac{2\cdot(dnum-1)\cdot(l+K+1)}{dnum}$ for ModUp as explained in \autoref{sec:lc_dist}. During ModDown, $2\cdot K$ polynomials would be communicated for BaseConversion.  ModDown can also be handled within each chiplet, eliminating the need for cross-chiplet data exchange if the result of Key Multiplication is duplicated across all chiplets, as depicted in the second figure of  \autoref{fig:exch_1}. This results in a one-time communication cost of $\frac{2\cdot(dnum-1)\cdot(l+1)}{dnum} + 2\cdot K$ polynomials per chiplet and requires additional storage for $2\cdot (K-1)$ polynomials in each chiplet.

Each chiplet would have to perform duplicate $K$ INTT operations and BaseConversion steps during ModDown while the $(l+1)$ NTT operations would be distributed across the chiplets. A key limitation of this technique is that not much computation can be done in parallel with the communication after ModUp. While the polynomial transfers can overlap with the NTT and Key Multiplication computations, the high throughput of NTT+KeyMult operations means this overlap does not achieve the desired decoupling effect. As a result, the potential for parallelism is limited, leading to delays. Additionally, this approach demands many chiplets for higher values of $dnum$. As the digits for the key switch decrease, many chiplets become idle, further reducing efficiency. This inefficiency in fully utilizing chiplets highlights a bottleneck in achieving optimal performance.

\begin{figure}[t]
    \centering
    \includegraphics[width=0.75\linewidth,trim={0.8cm 0.3cm 3cm 0.2cm},clip]{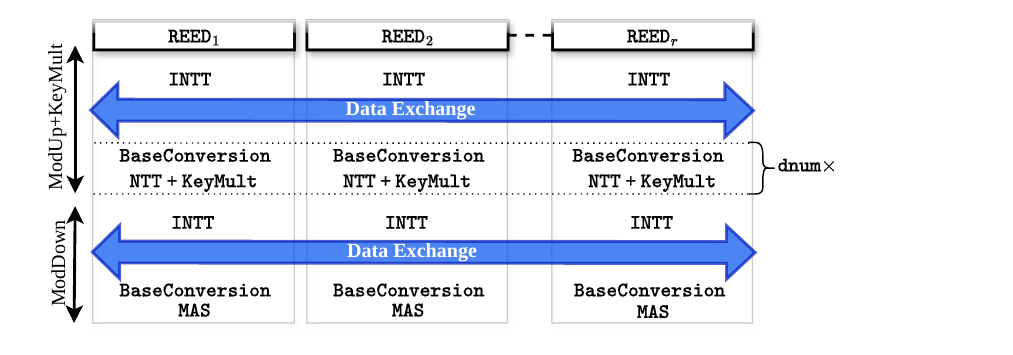}
    \caption{The Data and Task distribution when limbs are distributed alternately for KeySwitch computation. The Data Exchange is done in parallel with INTT, BaseConversion, and NTT+KeyMult of previous digits.}
    \label{fig:exch_2}
\end{figure}

In contrast, the alternate limb distribution technique distributes the task for each digit across a fixed number of chiplets, as illustrated in \autoref{fig:exch_2}. This means communication is only required during ModUp/ModDown, after which all the data needed for Key Multiplication and accumulation remains confined within a single chiplet. During ModUp, the chiplets broadcast the INTT results (similar to the case of $dnum=L+1$), which is efficient as only $l+1$ polynomials undergo INTT conversion. After ModUp, the number of polynomials increases to $dnum\cdot (l+K+1)$. Although communication is necessary for both ModUp and ModDown, our technique minimizes communication overhead since the number of polynomials broadcasted during ModUp ($l+1$) and ModDown ($2\cdot K$) is significantly lower compared to the previous method's $\frac{2\cdot(dnum-1)\cdot(l+1)}{dnum} + 2\cdot K$ for $dnum>2$. 

 We investigated the possibility of using data duplication to decouple INTT data transfer from NTT computation. However, we found that the overhead caused by duplicating data after each operation outweighed any potential communication benefits during ModUp and ModDown. Thus, we converge to an approach where the chiplets do not compute only $K$ INTT required for each digit computation during ModUp but all $l+1$ INTT operations, which they would require for computation corresponding to all the digits.

\subsection{Throughput Computation of KeySwitch for $dnum<L+1$}\label{sec:appendix_throughput}

In this section, we discuss how the throughput is obtained when $dnum=3$, in line with \autoref{sec:th_ks}, in the four ($=r$) chiplet setting of \papertitle. As discussed above we, perform all $l+1$ INTT at once, to bridge the computation communication gap. Thus, each chiplet performs at most $\frac{l+1}{r}$ INTT. For the first set of BaseConversion followed by Key Multiplication, only $\frac{l+1}{dnum}$ INTT results are needed. Since, these limbs are distributed across chiplets a communication of $\frac{(r-1)\cdot(l+1)}{dnum\cdot r}$ is required to every chiplet. This communication also works in ring manner, similar to the strategy proposed in~\autoref{sec:non_blocking_comm}.

 \begin{figure}[t]
    \centering
    \includegraphics[width=\linewidth,trim={0.6cm 0.2cm 2.2cm 0.2cm},clip]{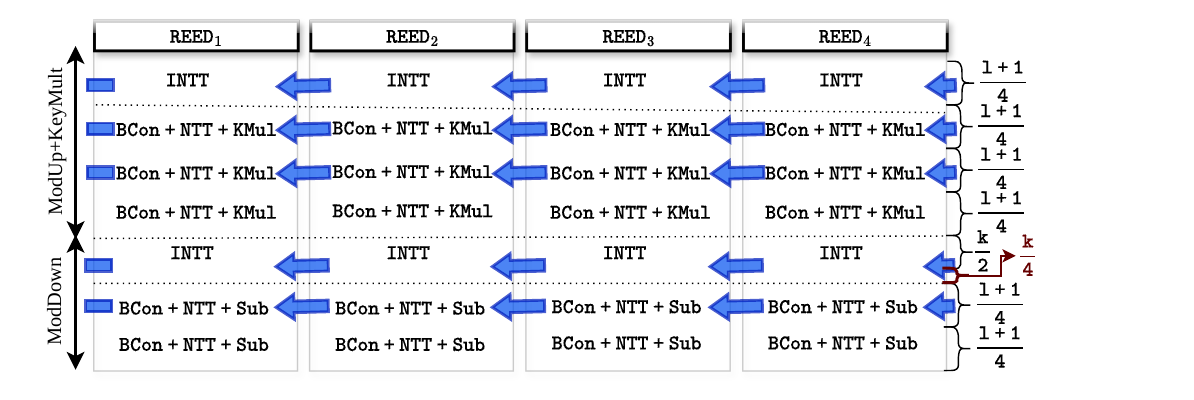}
    \caption{KeySwitch Flow showing computation throughput and communication overhead for ModUp, KeyMult and ModDown for $dnum=3$. }
    \label{fig:onechiplet}
\end{figure}

Note that we cannot start NTT before we have accumulated all the $\frac{l+1}{dnum}$ results as required for \autoref{algo:bconv_gen}. When $dnum=L+1$, this value is 1, hence we can start the NTT immediately on the INTT result, however, the same cannot be done for lower values of $dnum$. Thus, in the prior case, while a long communication window was available, here the communication window is limited. The first set of  $\frac{(r-1)\cdot(l+1)}{dnum\cdot r}$ INTT results have to be communicated while the chiplet is computing $\frac{l+1}{r}$ INTTs.  Each BaseConversion computes $l+1$ NTTs ($\frac{l+1}{r}$ NTTs), which gives $\frac{dnum}{r-1}\times$ larger communication window for the INTT results required for subsequent BaseConversions. Note that, for $dnum=3$ and $r=4$, its value is one, hence we only face computation overhead for the last polynomial as the C2C communication offers the same throughput as the configuration $1024\times 64$. However, when the C2C communication becomes slower, than this directly impacts the communication and hence the delay before the BaseConversions.

Now, the goal of computation is to prevent computation from becoming the bottleneck in this case. Thus, instead of sending plain INTT results, the results are multiplied with $\hat{p}^{-1}_i$ for Step 5~\cite{ARK}, in \autoref{algo:bconv_gen}. After this the chiplets only need to perform a MAS operation on all the INTT results for each new base during BaseConversion. We are only restricted by off-chip communication bandwidth, and therefore we wait until all the polynomials have been shared. The  polynomials in INTT form are in on-chip memory. The memory for each polynomial is split across various SRAMs. Therefore, one polynomial can be loaded at a higher throughput. The operations on these can be handled by an extra MAS unit for BaseConversion using $K\times$ ($K=8$ for $dnum=3$) more multipliers to offer the same throughput as NTT. Thus, the BaseConversion, followed by NTT, and Key Multiplication and accumulation works in a pipeline. The key multiplication with the pre-existing $K$ bases is done in parallel to their INTT conversion, as it does not require any pre-computation. Hence, the overhead of the ModUp+KeyMul is directly correlated to the computation time of the INTT polynomials $\frac{(l+1)}{r}$ and the $dnum$ BaseConversions $\frac{dnum\cdot(l+1)}{r}$, as shown in \autoref{fig:onechiplet}.

While, the communication could be completely decoupled for the ModUp+KeyMul step, the decoupling reduces for ModDown. Each chiplet computes the $\frac{K}{r}$ INTT results and requires communication of $\frac{(r-1)\cdot K}{r}$ INTT results before it can perform the BaseConversion requiring $\frac{(l+1)}{r}$ NTT  computations per chiplet and the subtraction in pipeline. Thus, the overall overhead of this step is determined by communication of $\frac{(r-1)\cdot K}{r}$ polynomials and followed by $\frac{(l+1)}{r}$ NTT computation. This is incurred twice, once for each ciphertext component. Instead of doing the ModDown sequentially for each ciphertext component, we follow the technique proposed earlier and perform all the INTT $\frac{2\cdot K}{r}$ INTT computations at once.  Thus the communication overhead for the first ModDown is reduced to $\frac{(r-3)\cdot K}{r}$. While the first ModDown is being computed the limbs for the next ModDown are communicated. Thus, the delay due to communication overhead of $\frac{(r-3)\cdot K}{r}$ is incurred only once, and the computation overhead is $\frac{2\cdot(l+1+K)}{r}$ NTT/INTT computations. 

Overall, this results in runtime complexity of $\frac{(2\cdot(l+1+K)+(dnum+1)\cdot(l+1)+(r-3)\cdot K)}{r}$. With $dnum=3$ and $r=4$, the communication overhead is ${\approx}{5}\%$, while INTT/NTT is being computed in parallel and MAS operations in pipeline. Substituting the values of $dnum=L+1$, shows how this reduces to bare minimum. Thus, NTT unit stays almost fully utilized (${\approx}{95}\%$) during the hybrid key-switch for $dnum=3$. We have kept our analysis generic, so that similar results can be derived for varying values of $dnum$ and $r$. This also shows how $r=4$, gives a sweet spot, and a higher value would result in higher communication delay, not only during ModDown but also during ModUp, for lower values of $dnum$. Note that, this $5\%$ communication delay can be completely bridged by interleaving the last BaseConversion+NTT+KeyMul step with the INTT required for ModDown, as no communication happens during this step. However, we leave this dataflow utilization for future works.

\bibliographystyle{alpha}
\bibliography{refs}

\end{document}